\documentclass[aps,pra,twocolumn,floats,showpacs,floatfix]{revtex4}
\pdfoutput=1 % within first 5 lines!
\usepackage{color}
\usepackage{bm}
\usepackage{latexsym}
\usepackage{amsmath}

\usepackage{graphicx}
\begin{document}

\renewcommand{\dag}[1]{#1^{\dagger}}
\newcommand{\stext}[1]{_{\mbox{\tiny#1}}}
\newcommand{\tbox}[1]{\mbox{\tiny#1}}
\newcommand{\cl}[1]{#1_{\tbox{cl}}}
\newcommand{\half}{\small\frac{1}{2}}
\newcommand{\ecl}{\delta E_{\tbox{cl}}}
\newcommand{\mbf}[1]{\mathbf{#1}}
\newcommand{\mcal}[1]{\mathcal{#1}}
\newcommand{\szp}{\;.}
\newcommand{\szk}{\:,}
\newcommand{\sinc}{\mbox{sinc}}
\newcommand{\dd}[1]{\mbox{d}#1\,}
\newcommand{\ddp}[1]{\partial#1}
\newcommand{\e}[1]{\, e^{#1}}
\newcommand{\abs}[1]{\left|#1\right|}
\newcommand{\com}[2]{\left[#1,#2\right]}
\newcommand{\bra}[1]{\left\langle #1\right|}
\newcommand{\ket}[1]{\left|#1\right\rangle }
\newcommand{\scap}[2]{\langle#1|#2\rangle}
\newcommand{\braket}[3]{\langle#1|#2|#3\rangle}
\newcommand{\tket}[1]{|#1\rangle}
\newcommand{\tbra}[1]{\langle#1|}
\newcommand{\tscap}[2]{\langle#1|#2\rangle}
\newcommand{\tbraket}[3]{\langle#1|#2|#3\rangle}
\newcommand{\id}{1\hspace*{-1ex}\rule{0.15ex}{1.3ex}\hspace*{1.1ex}}
\newcommand{\cc}[1]{#1^{*}}
\newcommand{\dog}[1]{#1^{\dagger}}
\newcommand{\grad}{\vec{\nabla}}
\newcommand{\hide}[1]{}
\newcommand{\revision}[1]{{\color{red}{#1}}}

%%%%%%%%   TITLE    %%%%%%%%%%%%%%%%%%%%%%%%%%%%

\title{Wavepacket dynamics in energy space of a chaotic trimeric Bose-Hubbard system}

\author{
Moritz Hiller$^{1}$, Tsampikos Kottos$^{2}$ and Theo Geisel$^{3}$
}

\affiliation{
$^1$Physikalisches Institut, Albert-Ludwigs-Universit\"at, Hermann-Herder-Str.~3, D-79104 Freiburg, Germany\\
$^2$Department of Physics, Wesleyan University, Middletown, Connecticut 06459, USA\\
$^3$MPI for Dynamics and Self-Organization, Bunsenstra\ss e 10, D-37073 G\"ottingen, Germany
}

\begin{abstract}
We study the energy redistribution of interacting bosons in a {ring-shaped} quantum trimer as
the coupling  strength between neighboring sites of the corresponding Bose-Hubbard Hamiltonian
undergoes  a sudden change $\delta k$. Our analysis is based on a three-fold approach combining
linear response theory calculations as well as semiclassical and random matrix theory considerations. 
The $\delta k$-borders of applicability of each of these methods are identified by direct
comparison with the exact quantum mechanical results. We find that while the variance of the 
evolving quantum distribution shows a remarkable quantum-classical correspondence (QCC) for 
all $\delta k$-values, other moments exhibit this QCC only in the non-perturbative $\delta k$-regime. 
\end{abstract}
\maketitle

%%%%%%%%%%     INTRODUCTION    %%%%%%%%%%%%%%%%%%%

\section{Introduction}

Understanding the intricate behavior of bosonic many-body systems has been a major challenge for
leading research groups over the last years. Without doubt, the theoretical interest was strongly
enhanced by recent experimental achievements in handling (ultra-)cold quantum gases: Namely,
since the celebrated realization of atomic Bose-Einstein condensates (BEC) in periodic optical
lattices (OL) \cite{AK98,JBCGZ98,CBFMMTSI01,OTFSYK01} and the creation of {}``atom chips'' \cite{FKCHMS00,HHHR01,OFSGZ01,Reich02} we have versatile tools at hand which allow for an
unprecedented degree of precision as far as manipulation and measurement of the atomic cloud
is concerned. While this has led on the one hand to novel, concrete applications of quantum
mechanics like e.g. atom interferometers \cite{ACFAHS02,SHAWGBSK05,WAB_etal_05} and lasers
\cite{MAKDWK97,ATMDKK97,AK98, HDKWHRP99,OTFSYK01} it also enabled us to investigate
complex solid-state phenomena, such as the Mott-Insulator to superfluid transition \cite{GMEHB02}
or the Josephson effect \cite{CBFMMTSI01}. 

Beside these advances, our understanding of bosonic many-body systems is still very limited once we consider
an (external) driving: In the framework of BECs in OLs this can be, e.g., a modulation of the potential height
or a tilting of the lattice. Due to the time dependence of the driving parameter, the energy of the system
is not a constant of motion. On the contrary, the system experiences ÔÔtransitionsÕÕ between energy
levels and therefore absorbs energy. This irreversible loss of energy is know as dissipation \cite{Wei98,
Wil99_B, Cohe00, HCGK06}. The classical dissipation mechanism is by now well understood \cite{Wei98}
while quantum dissipation still poses some challenges. In order to get a better insight into the problem the
main task is to build a theory for the time-evolving energy distribution. Apart from being of fundamental
interest, such a theory will also shed a new light on recent experiments with BECs amplitude-driven OLs \cite{SMSKE04,SSMKE04} .
It has been pointed out that the energy absorption rate measurements can be used to probe the many-body
excitations of the system \cite{KIGHS06,RBPWC05,ICHG06,BASD05,Lund04b}.

In the present work, we approach the problem of quantum dissipation by studying the quantum dynamics of
interacting bosons on a ring-shaped lattice consisting of three sites (trimer). Specifically, we will analyze the
system's response to a rectangular pulse of finite duration $t$ that perturbs the coupling
$k_0\rightarrow k=k_0+\delta k$ between adjacent sites. In the framework of OLs this corresponds to a sudden
change in the intensity of the laser field and is readily achieved in the experiment \cite{SSMKE04,SMSKE04}.
The associated dynamical scenario known as \emph{wavepacket dynamics} \cite{Cohe00,CIK00,KC01,HKG06}
is one of the most basic non-trivial evolution schemes. Its analysis will pave the way to understand more
demanding evolution scenarios and ultimately the response of interacting bosons under persistent driving. 

The minimal quantum model that describes interacting bosons on a lattice $M$ wells
is the Bose-Hubbard Hamiltonian (BHH), which incorporates the competition between kinetic and
interaction energy of the bosonic system. The BHH is based on a $M$-mode approximation and
hence its validity is subject specific conditions discussed in Refs.~\cite{MCWW97,JBCGZ98,MO06,BDZ08}
(see also Section \ref{cha2-sec:BHH}). As far as the BHH is concerned, the two-site system (dimer)
has been analyzed thoroughly from both the classical (mean-field) \cite{FPZ00,ELS85,TK88} and the purely
quantum viewpoint \cite{FPZ00,BES90,KBK03} and many exciting results were found including
their experimental realization \cite{AGFHCO05}. 

As a matter of fact, the dimer is integrable since the BHH has two conserved quantities, the energy and the 
number of bosons. The addition of a further site {--yielding either a linear chain (open bc)
or a ring (periodic bc)--} is sufficient to make the resulting system (trimer) non-integrable and thus
leads to (classically) chaotic behavior. {Here we consider a three-site ring 
 \footnote{{We expect qualitatively similar results for a linear chain configuration.}}
which can be experimentally realized} using optical lattice or micro trap technology \cite{OFSGZ01,HHHR01,
Reich02,KSHBESBFFS03}. For example, an optical potential in a ring configuration can be achieved 
by letting a plane wave interfere with the so-called Laguerre-Gauss laser modes 
as described in \cite{AOC05,BF07}. Another possibility to experimentally create a three-site 
ring BEC trap \footnote{We thank N. Davidson for pointing this out to us.} is given by a combination 
of the methods described in Refs.~\cite{SWTNA07,LLSS07}: In the experiment of \cite{SWTNA07}
a trapping potential is partitioned into three sections by a central repulsive barrier created 
with blue-detuned laser light that is shaped to segment the harmonic oscillator potential well 
into three local potential minima. For better optical resolution (up to $1.2 \mu m$), and control
of the coupling between the three condensates one can substitute the detuned laser source of 
Ref.~\cite{SWTNA07} with the one used in \cite{LLSS07}.

The motivation to study the quantum trimer is twofold: That is, while remaining simple enough to 
allow for a thorough analytical study, it displays a whole new class of complex behaviors which 
are typical for longer lattices consisting of many sites. The trimer has been studied quite extensively
in the classical (mean-field) regime 
\cite{ETT95,FP02,CPC00}. Less attention was paid to the analysis of the quantum trimer \cite{FFS89,
CFFCSS90,NHMM01,HKG06,BHK07,PF06,KB03}. As a matter of fact, the majority of these studies is focused on 
the statistical properties of levels \cite{FFS89,CFFCSS90,KB04} while recently an analysis of 
the shape of eigenstates was performed in Ref.~\cite{HKG06}. However, the knowledge of spectral 
and wavefunction statistics is not enough if one wants to predict the dynamical behavior of a 
system.

In our study we combine three theoretical approaches: On the one hand, we will use linear
response theory (LRT) which constitutes the leading framework for the analysis of driven
systems \cite{Cohe00}. On the other hand, we employ an improved random matrix theory 
(IRMT) modeling. Although random matrix theory (RMT) was proven to be a powerful tool
in describing \emph{stationary} properties (like level statistics \cite{KB04,CFFCSS90} and
eigenfunctions \cite{HKG06}), its applicability to the description of wavepacket dynamics is
not obvious \cite{KC01,HCGK06}. The latter involves not only the knowledge of the statistical
properties of the two quantities mentioned above but also the specific correlations
between them. Finally, we will investigate the validity of semiclassical methods to describe the
quantum evolution. Our analysis indicates that some moments of the evolving energy distribution
show a remarkable level of quantum-classical correspondence (QCC) \cite{CIK00,KC01,HCGK06} 
while others are strongly dominated by quantum interference phenomena.

The structure of this paper is as follows: in the next section, we introduce 
the Bose-Hubbard Hamiltonian that mathematically describes a quantum three-site ring-lattice.
We identify its classical limit, leading to the discrete nonlinear Schr\"odinger equation and
derive the classical equations of motion. In Section \ref{sec:object} we discuss the notion of
wavepacket dynamics and introduce the observables studied in the rest of the paper.  We begin
our analysis with the statistical properties of the spectrum and of the matrix elements of the BHH
(Section~\ref{sec:statistics}). This study allows us to introduce an IRMT modeling which is 
presented in Subsection \ref{cha3-sec:RMT}. In Section \ref{cha3-sec:P_nm_of_BHH} we extend 
our previous analysis on the parametric evolution of the eigenstates of the BHH \cite{HKG06}
by comparing the actual quantum mechanical calculations with the results of the IRMT 
modeling. We introduce the concept of parametric
regimes \cite{HKG06}  and show how it can be applied to analyze the parametric evolution of the
local density of states (LDoS) \cite{CK01,HCGK06,HKG06}. We then turn to the dynamics of the
BHH (Section \ref{cha4-sec:WP-dyn}) and extend the notion of regimes to the wavepacket dynamics
scenario. The predictions of LRT, IRMT modeling and semiclassics are compared with the exact
quantum mechanical calculations for the trimeric BHH model. We find that the energy spreading
$\delta E(t)$  shows a remarkable quantum-classical correspondence which is independent of the
perturbation strength $\delta k$. In contrast, other observables are sensitive to quantum interference
phenomena and reveal QCC only in the semiclassical regime. The latter can be identified with the
non-perturbative limit associated with perturbations $\delta k > \delta k_{\rm prt}$. Section
\ref{sec:conclusions} summarizes our findings.

%%%%%%%%%%%%%%    BHH-DNLS   %%%%%%%%%%%%%%%%%%%%%%
\section{The Bose-Hubbard Hamiltonian\label{cha2-sec:BHH}}

The mathematical model that describes interacting bosons in a (three-site) lattice is the
Bose-Hubbard Hamiltonian, which in second quantization reads
\begin{equation}
   \hat{H}=
%\sum_{i=1}^{3}\epsilon_{i} {\hat n}_i +
   \frac{U}{2}\sum_{i=1}^{3}{\hat n}_i ({\hat n}_i-1) -
   k\sum_{i\neq j}\hat{b}_{i}^{\dagger}\hat{b}_{j};\quad \hbar=1
   \label{eq:BHH}
\end{equation}
{Here we consider a three-site ring configuration which} is experimentally feasible with current optical methods where,
for example, the trapping potential is created by letting a plane wave interfere with the so-called Laguerre-Gauss
 laser modes as described in \cite{AOC05}. The operators ${\hat n}_i=\hat{b}_i^{\dagger}
\hat{b}_i$ count the number of bosons at site $i$. The annihilation and creation operators 
$\hat{b}_i$ and $\hat{b}_i^{\dagger}$ obey the canonical commutation relations $[\hat{b}_i,
\hat{b}_j^{\dagger}]=\delta_{i,j}$. In the BEC 
framework, $k=k_0+\delta k$, is the coupling strength between adjacent sites $i,j$, and can 
be controlled experimentally (in the context of optical lattices this can be achieved by 
adjusting the intensity of the laser beams that create the trimeric lattice), while $U=4 
\pi\hbar^2a_s/m V_{\rm eff}$ describes the interaction between two atoms on a single site ($m$ is the 
atomic mass, $a_s$ is the $s$-wave scattering length of atoms which can be either positive 
or negative, and $V_{\rm eff}$ is the effective mode volume). It is interesting to note that the
BHH also appears in the context of molecular physics where \cite{JGBSF05,BES90} $k$
represents the electromagnetic and mechanical coupling between bonds of adjacent
molecules $i,j$, while $U$ represents the anharmonic softening of the bonds under extension.

{The Bose-Hubbard model for $M$ sites is based on a $M$-mode approximation \cite{MCWW97}
(in the limit of long lattices this corresponds to a single (lowest) band approximation of the OL \cite{JBCGZ98}).
This assumption holds provided that the chemical potential,
the kinetic energy and the interaction energy are too low to excite states in the higher single-well modes
(higher Bloch bands accordingly). Therefore, the lattice must be very deep \cite{MCWW97,DGPS99,MO06} inducing
large band gaps. Furthermore, the interaction energy has to be smaller than the single
particle ground state energy, so as to not considerably modify the single particle wavefunction.
A Gaussian approximation of the wavefunction together with a standard harmonic trap of
size $10\mu m$ and a scattering length $a_s = 5nm$ indicates that the BHH model is valid
for up to several hundred bosons per trap \cite{MCWW97}. 
}

 Hamiltonian (\ref{eq:BHH}) 
has two constants of motion, namely the energy $E$ and the number of particles $N= \sum_{i=
1}^3n_{i}$. Having $N=const.$ implies a finite Hilbert-space of dimension ${\cal N}=(N+2)
(N+1)/2$ \cite{FFS89,BES90} which can be further reduced by taking into account the threefold
permutation symmetry of the model \cite{HKG06}. 
   
%%%%%%%%%  Classical limit   %%%%%%%%%%%%%%%%%%%

For large particle numbers $N\gg1$ one can adopt a semiclassical approach for Hamiltonian
(\ref{eq:BHH}). Formally, this can be seen if we define rescaled creation and annihilation operators
$\hat{c}_i =\hat{b}_i/\sqrt{N}$. The corresponding commutators $[\hat{c}_i,\hat{c}_j^{\dagger}]=\delta_{ij}/N$
vanish  for $N\gg1$ and therefore one can treat the rescaled operators as c-numbers. Using the Heisenberg 
relations $\hat{c}_i\rightarrow \sqrt{I_i}\exp^{i\varphi_{i}}$ ($\varphi_i$ is an angle and $I_i$ is the associated
action \footnote{The quantum mechanical conservation of the particle number $N$ translates into conservation
of total action $I=\sum_{i}I_{i}$.} ), we obtain the classical Hamiltonian ${\cal H}$ 
\begin{equation}
   {\tilde {\cal H}} = \frac{\cal H}{N{\tilde U}} =\frac{1}{2}\sum_{i=1}^{3}I_{i}^{2}-
   \lambda\sum_{i\neq j}{\sqrt {I_{i}I_{j}}}\exp^{i(\varphi_j-\varphi_i)}\, ,
   \label{eq:H-DNLS}
\end{equation}
where ${\tilde U}=NU$ is the rescaled on-site interaction.

The dynamics is obtained from (\ref{eq:H-DNLS}) using the canonical equations $dI_i/d{\tilde t} =
-\partial {\tilde {\cal H}}/\partial \varphi_i$ and $d\varphi_i/d{\tilde t} = \partial {\tilde {\cal H}}/\partial I_i$.
Here ${\tilde t}= {\tilde U} \cdot t$ is the rescaled time. The classical dynamics depends both on the
scaled energy $\tilde{E}=E/\tilde{U}N$ and the dimensionless parameter $\lambda=k/\tilde{U}$ \cite{SLE85,
TK88,FFS89,FP02,NHMM01}. For $\lambda\rightarrow0$ the interaction term dominates and the system behaves 
as a set of uncoupled sites (also known as the \emph{local-mode} picture \cite{BES90}) while in the 
opposite limit of  $\lambda\rightarrow\infty$, the kinetic term is the dominant one (\emph{normal-mode} 
picture \cite{SLE85,ELS85,WEHMS93}). In both limits the motion is integrable while for intermediate 
values of $\lambda$ the trimeric BHH  (\ref{eq:BHH}) has a chaotic component \cite{CFFCSS90}. We point out 
that the classical limit is approached by keeping $\lambda$ and $\tilde{U}$ constant while $N
\rightarrow\infty$ \cite{HKG06}. This is crucial in order to keep the underlying classical motion 
unaffected.

%%%%%%%%%%%%%%     OBJECT OF THE STUDY   %%%%%%%%%%%%%%%
\section{Preliminary considerations and object of the study\label{sec:object}}

In this paper we study the trimeric BHH model (\ref{eq:BHH}) as a control parameter, the coupling 
strength between lattice sites is changed i.e. $k_{0}\rightarrow k_{0}+\delta k$. In our analysis, 
we therefore consider 
\begin{equation}
  \hat{H}=\hat{H}_{0}-\delta k(t)\hat{B}\szk
  \label{cha3-eq:linear-Hamiltonian}
\end{equation}
where the perturbation operator $\hat{B}$ is
\begin{equation}
  \hat{B}=\sum_{\langle i,j\rangle}\hat{b}_{i}^{\dagger}\hat{b}_{j}\szk
  \label{cha3-eq:B-operator}
\end{equation}
and the unperturbed Hamiltonian $\hat{H}_{0}$ is given by Eq.~(\ref{eq:BHH})
with $k=k_{0}$. Quantum mechanically, we work in the $\hat{H}_{0}$
eigenbasis. In this basis $\hat{H}_{0}$ becomes diagonal, i.e., $\mbf{E}_{0}=E_{m}^{(0)}\delta_{mn}$
where $\{ E_{m}^{(0)}\}$ are the ordered eigenvalues and we can write
\begin{equation}
  \mbf{H}=\mbf{E}_{0}-\delta k\mbf{B}\szp
  \label{cha3-eq:BHH-hamilton_matrix}
\end{equation}
 
Throughout this work we \emph{always} assume that the perturbed Hamiltonian
$\mcal{H}(k)$ as well as the unperturbed Hamiltonian $\mcal{H}(k_{0})$
generate classical dynamics of the \emph{same nature}, i.e.,
that the perturbation $\delta k=k-k_{0}$ is \emph{classically small},
$\delta k<\delta k_{\tbox{cl}}$ (see beginning of the next section for the definition
of $\delta k_{\tbox{cl}}$). This assures the applicability of
classical linear response theory (LRT). Note, however, that this assumption
is not sufficient to guarantee the validity of quantum mechanical
linear response theory. Our aim is to identify novel quantum mechanical effects that
influence the classical LRT results as the perturbation $\delta k$ increases.
At the same time, we address the implications of classically chaotic
dynamics for the trimeric BHH, and the route to quantum-classical correspondence
in the framework of wavepacket dynamics.

\begin{figure}[!t]
\includegraphics[%
  width=0.6\columnwidth,
  keepaspectratio]{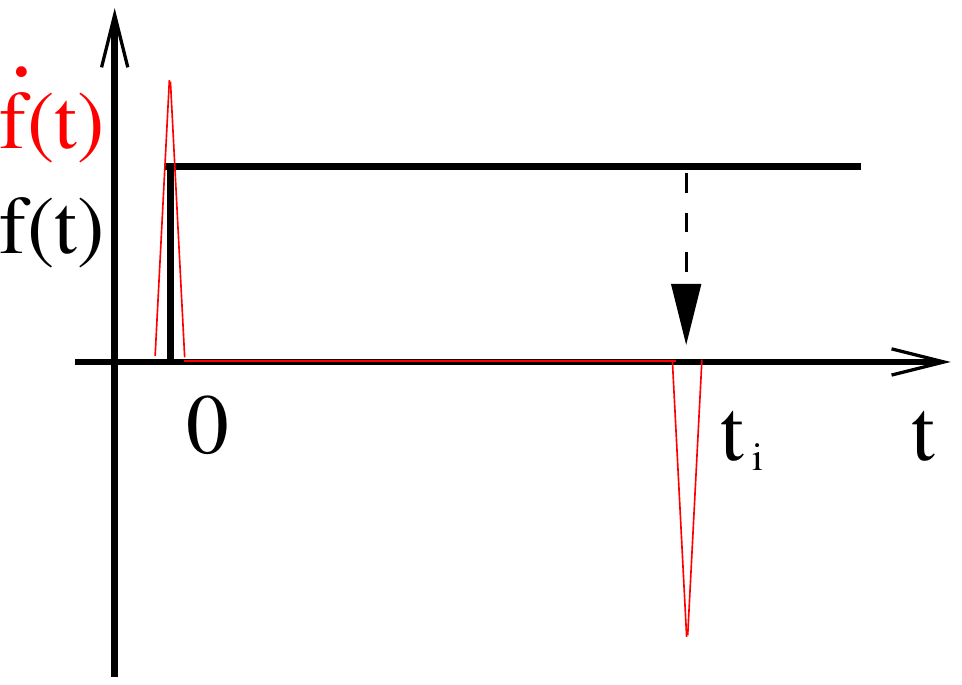}

\caption{\label{cha4-fig:f_t_sketch} (Color online) Scheme of the wavepacket dynamics scenario:
the perturbation is a rectangular pulse of duration $t_{i}$ at which
the measurement is done. The function $f(t)$ represents the rescaled
time-dependence of the perturbation $\delta k(t)=\delta k\times f(t)$
(black line) while the red line indicates its time derivative $\dot{f}(t)$.}
\end{figure}

For later purposes it is convenient to write the perturbation as $\delta k(t)=\delta k\times f(t)$
where $\delta k$ controls the {}``strength of the perturbation'' while $f(t)$ is the scaled time 
dependence (note that if we had $f(t)\propto t$, i.e. persistent driving, then $\delta k$ would be 
the {}``rate'' of the driving). Although our focus will be on the wavepacket dynamics scenario where
the perturbation is a rectangular pulse of strength $\delta k$ and duration $t$ --see 
Fig.~\ref{cha4-fig:f_t_sketch} for a sketch of the resulting step function $f(t)$ with $k(t)=k(0)$--
we expect that the results presented here will shed some light to the response of BHHs in the 
presence of more demanding driving scenarios. 

\begin{figure*}[t]
\hfill{}\includegraphics[%
  width=1.0\textwidth,
  keepaspectratio]{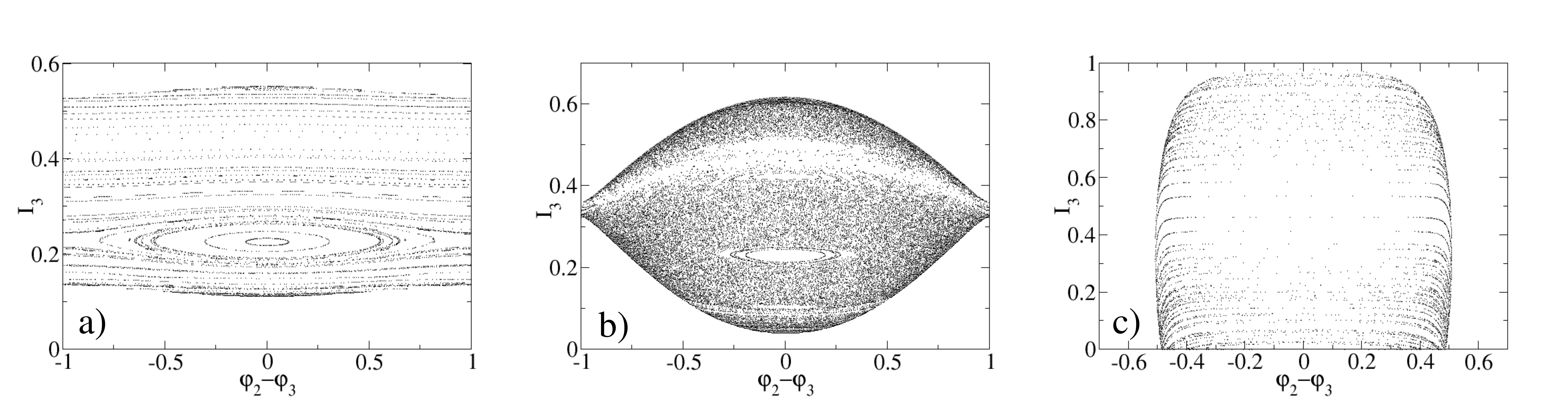}\hfill{}

\caption{\label{cha2-fig:BHH-poincare} Poincar\'e sections of the phase space
belonging to the classical trimer for $N=1$ and different parameter values
a) $\lambda=0.005$, b)$\lambda=0.05$, c) $\lambda=2$. On the $y$-axis we plot
the action $I_{3}$ while on the $x$-axis the difference $\varphi_{2}-\varphi_{3}$ (in
units of $\pi$) is plotted. The Poincar\'e section corresponds to the plane $\varphi_1=\varphi_3$
and $\dot{\varphi_1}>\dot{\varphi_2}$ of the energy surface $\tilde{E}=0.2$.}
\end{figure*}

%%%%%  measures of the profile  %%%%%
\subsection{Measures of the evolving distribution $P_{t}(n|n_{0})$\label{sec:measures}}

In this subsection we discuss a number of observables that will allow us to 
quantify the response of the system and the spreading of the energy distribution. 

We consider an initial micro-canonical preparation described by an eigenstate
$\tket{n_{0}}$ of the unperturbed Hamiltonian $\hat{H}(k(0))$.
Given the driving scenario $k(t)$, it is most natural to analyze
the evolution of the probability distribution 
\begin{equation}
P_{t}(n|n_{0})=|\langle n|\hat{U}(t)|n_{0}\rangle|^{2}\szk
\label{cha4-eq:Pnm_t}
\end{equation}
where
\begin{equation}
  \hat{U}(t)=\hat{T}\exp[-\frac{i}{\hbar}\int_{0}^{t}\dd{t'}\hat{H}(k(t')]
  \label{cha4-eq:U_operator}
\end{equation}
is the time-ordered evolution operator and $\hat{H}[k(t)]\tket{n[k(t)]}=E_{n}[k(t)]\tket{n[k(t)]}$.
By convention we order the states by their energy. Hence we can regard $P_{t}(n|n_{0})$ as a function 
of $r=n-n_{0}$, and average over the initial preparation (around some {\it classically} small energy
window), so as to get a smooth distribution $P_{t}(r)$.

To capture various aspects of the evolving probability distribution $P_{t}(n|n_{0})$
we introduce here the survival probability defined as
\begin{equation}
  P(t)=|\langle n_{0}|\hat{U}(t)|n_{0}\rangle|^{2}=P_{t}(n_{0}|n_{0})\szk
  \label{cha4-eq:P_SR}
\end{equation}
and the energy spreading 
\begin{equation}
  \delta E(t)=\sqrt{\sum_{n}P_{t}(n|n_{0})(E_{n}-E_{n_{0}})^{2}}\szk
  \label{cha4-eq:dE_t}
\end{equation}
which probes the tails of the evolving distribution. Yet, the evolution of $P_t(n|n_0)$ is 
not completely captured by any of these measures: As we will see in Section~\ref{cha4-sec:WP-dyn}
the wavefunctions can develop a {}``core'' which is a result of a non-perturbative mixing 
of levels \cite{HKG06}. We therefore define an operative measure that reflects the creation 
of the {}``core'', as the width $\delta E_{\tbox{core}}$ which contains $50\%$ of the probability: 
\begin{equation}
  \delta E_{\tbox{core}}(t)\,\,=\,\,[n_{75\%}-n_{25\%}]\Delta\szp
  \label{cha4-eq:dE_core_t}
\end{equation}
 Here, $\Delta$ is the mean level spacing and $n_{q}$ is determined
through the equation $\sum_{n}P_{t}(n|n_{0})=q$. 

%%%%%%%%%%    SPECTRA  AND  band profile  %%%%%%%%%%%%%%%%%%%%%%%%%%%%%%%%%%%%%%%%%%%%%
\section{Statistical properties of the trimeric BHH: spectra and band profile \label{sec:statistics}}

The dynamical properties of the classical trimer were thoroughly investigated in a number of papers 
\cite{ETT95,FP02,CPC00}. It was found that for intermediate values of the control parameter $\lambda$, 
the system exhibits (predominantly) chaotic dynamics. Some representative Poincar\'e sections (corresponding
to the plane $\varphi_1=\varphi_3$ and $\dot{\varphi_1}>\dot{\varphi_2}$ of the energy surface 
$\tilde{E}=0.2$ of Hamiltonian (\ref{eq:H-DNLS})) of the phase space are reported in 
Fig.~\ref{cha2-fig:BHH-poincare}. As $\lambda$ decreases, one can clearly see the transition from integrability to chaotic 
dynamics and back to integrability. We determine the regime of predominantly
chaotic motion based on the nature of the phase space and the power spectrum $\tilde{C}(\tilde{\omega})$
of the classical perturbation operator (the latter is discussed in detail in Subsection~\ref{cha3-sec:bandprofile}).
While regular motion results in isolated peaks in $\tilde{C}(\tilde{\omega})$, a continuous (but possibly structured)
power spectrum indicates chaoticity. Accordingly, the classical smallness condition $\delta k\ll\delta k_{\rm cl}$
can be operatively defined as the perturbation strength that leaves $\tilde{C}(\tilde{\omega})$ unaffected.
We have found that for $0.04<\lambda=k/\tilde{U}<0.2$ and an energy interval $\tilde{H} \approx0.26 \pm0.02$ the motion
is predominantly chaotic. Choosing our parameter values to be $k_0=15$ and $\tilde{U}=280$ we find $\delta k_{\rm cl}\approx20$. 

In the following we will concentrate on the above mentioned range of $\lambda$ values for which chaotic
dynamics is observed. The main question we will address is: What are the signatures of classical chaos
in various statistical quantities upon quantization? As we shall see in the following subsections, chaos 
manifests itself mainly in two quantities; the spectral statistics of the eigenvalues $\{E_m^{(0)}\}$
and the averaged profile $\langle|\mbf{B}_{mn}|^{2}\rangle$ of the perturbation operator. While the
statistical properties of the levels have attracted some attention in the past \cite{FFS89,Chef96}, the traces of chaotic 
dynamics in the shape of the perturbation operator $\langle|\mbf{B}_{mn}|^{2} \rangle$ and the 
statistical properties of its matrix elements were left unexplored. In the next subsections we 
will address these issues in detail and propose an improved random matrix theory modeling 
which takes our statistical findings into consideration.

\begin{figure}
\includegraphics[%
  width=1.0\columnwidth,
  keepaspectratio]{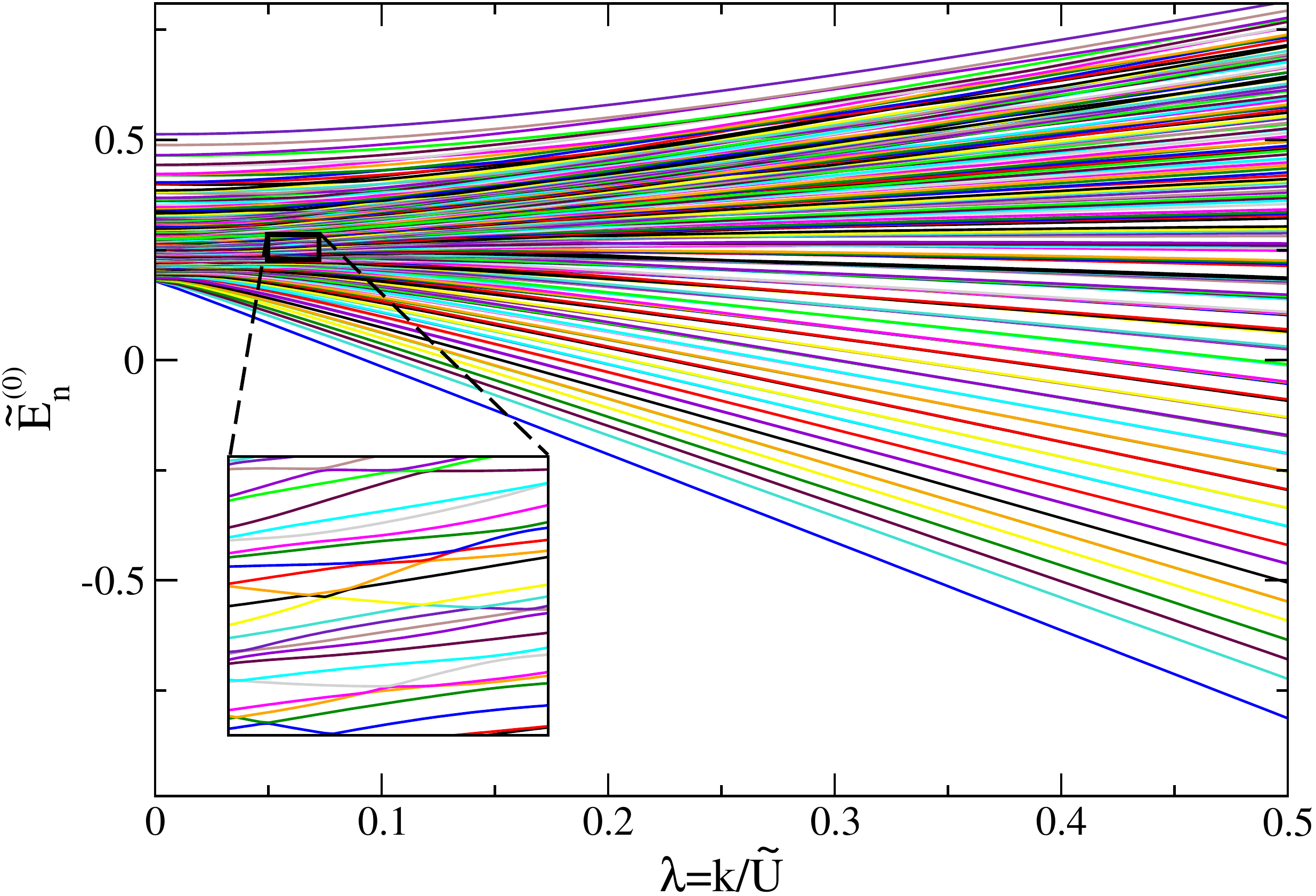}

\caption{\label{cha3-fig:pomodoro}  (Color online) Parametric evolution of the eigenvalues
$\tilde{E}_{n}^{(0)}$as a function of the parameter $\lambda$. The
number of bosons is $N=40$ and the effective interaction strength
is $\tilde{U}=280$. In the main figure the entire spectrum is plotted
while the inset is a magnification of the small box. One observes
a qualitative change in the spectrum as $\lambda$ is changed. See
text for details.}

\end{figure}

%%%%  Energy levels  %%%%
\subsection{Energy levels\label{cha3-sec:Energy-levels}}
\begin{figure*}[t!]
\includegraphics[%
  width=1.0\textwidth,
  keepaspectratio]{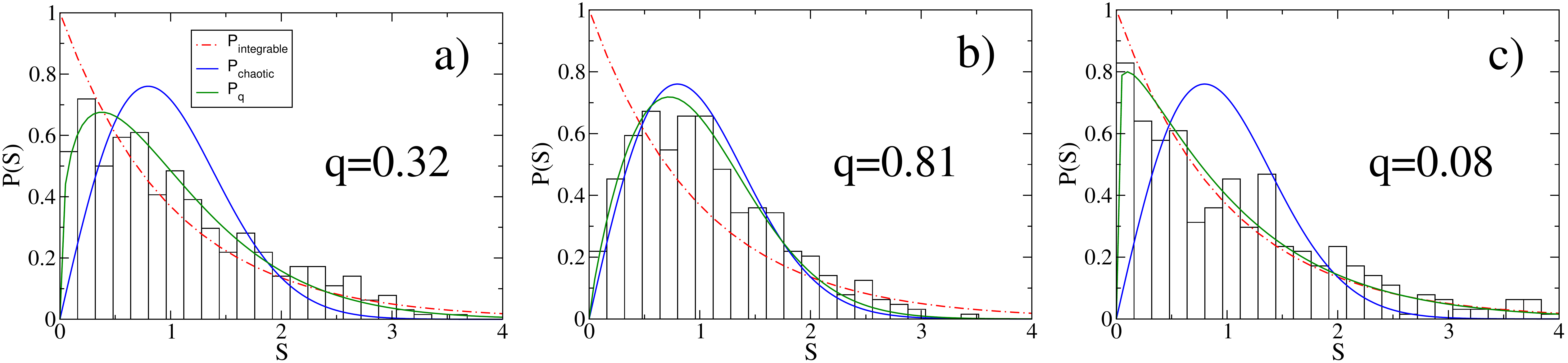}
\caption{\label{cha3-fig:LSS}  (Color online) The level spacing distribution ${\mathcal P}(S)$ of the
BHH trimer for three representative values of the dimensionless ratio
$\lambda=k/\tilde{U}$ which controls the underlying classical dynamics:
a) $\lambda=0.025\,(k=7)$, b) $\lambda=0.05\,(k=14.5)$, and c) $\lambda=0.35\,(k=100)$.
The red dash-dotted line corresponds to the Poissonian distribution
(\ref{cha3-eq:poisson}) which is expected for integrable systems,
the solid blue line corresponds to the Wigner surmise (\ref{cha3-eq:GOE})
(chaotic systems) while the solid green line represents the fitted
Brody distribution (\ref{cha3-eq:brody}). In Figure \ref{cha3-fig:LSS_brody}
we report the fitted Brody parameter $q$ for various values of $\lambda$.
The System corresponds to $N=230$ bosons and $\tilde{U}=280$. The
histograms include the $400$ relevant levels around $\tilde{E}=0.26$.}
\end{figure*}

In Fig.~\ref{cha3-fig:pomodoro} we plot the parametric evolution of the eigenvalues 
$\tilde{E}_{n}^{(0)}$ as a function of $\lambda$ for fixed effective interaction strength 
$\tilde{U}=280$. From Fig.~\ref{cha3-fig:pomodoro} one observes that the spectrum
becomes rather regular for very large $\lambda$. Indeed, for $\lambda\rightarrow\infty$
a transformation to the normal modes of the system diagonalizes the
Hamiltonian and yields an equidistant spacing of the eigenvalues \cite{FFS89}. In the
local-mode limit, i.e. $\lambda\rightarrow0$, the eigenvalues of $\hat{H}_{0}$ are obtained
immediately from (\ref{eq:BHH}) and are partly degenerate \footnote{We note that these
are 'accidental' degeneracies. In contrast, systematic degeneracies resulting from the symmetry of the
model are eliminated by restricting the calculations to the symmetric subspace \cite{FFS89}. See also
the following note.}.
However, in an intermediate $\lambda$-regime one observes a different behavior, namely
irregular evolution and level repulsion (see inset). This is a manifestation of the classically chaotic
behavior \cite{Haa00,Sto99}.

In order to establish this statement we turn to the statistical properties of the spectra. In 
particular, we will study the level spacing distribution ${\mathcal P}(S)$  \cite{FFS89,CFFCSS90,
Chef96,KB04} where
\begin{equation}
S_{n}=\frac{E_{n+1}-E_{n}}{\Delta}
\label{cha3-eq:Level-spacing}
\end{equation}
are the spacings of two consecutive energy levels which are \emph{unfolded} with respect to the 
local mean level spacing $\Delta$. The level spacing distribution represents one of the most popular 
measures used in quantum chaos studies \cite{Sto99,Haa00}. It turns out that the sub-$\hbar$ 
statistical features of the energy spectrum of chaotic systems are {}``universal'', and obey the 
RMT predictions \cite{Wign55,Wign57}. In contrast, non-universal, i.e. system specific, features 
are reflected only in the large scale properties of the spectrum and constitute the fingerprints 
of the underlying classical dynamics.

The mean level spacing $\Delta$ can be estimated from the fact that $\mathcal{N} \propto N^{2}$ 
levels span an energy window $\Delta E \propto \tilde{U}N\times\tilde{E}$, around some specific 
energy $\tilde{E}$ (see Eq.~(\ref{eq:BHH})). Our considerations indicate the scaling relation
\begin{equation}
\Delta\approx1.5\,\,\frac{\tilde{U}}{N}\szk
\label{cha3-eq:Delta_exact}
\end{equation}
where the proportionality factor was found by a direct fit of our spectral data in the energy 
window around ${\tilde E}=0.26$ \cite{HKG06}.

For chaotic systems the level spacing distribution ${\mathcal P}(S)$ follows the so-called 
\emph{Wigner} \emph{surmise} \cite{BGS84,Sto99}
\begin{equation}
  {\mathcal P}_{\tbox{ch}}(S)=\frac{\pi}{2}\, S\, e^{-\frac{\pi}{4}\, S^{2}}\szk
  \label{cha3-eq:GOE}
\end{equation}
indicating that there is a linear repulsion between nearby levels. Instead, for generic integrable 
systems there is no correlation between the eigenvalues and the distribution ${\mathcal P}(S)$ is Poissonian
\begin{equation}
  {\mathcal P}_{\tbox{int}}(S)=e^{-S}\szp
  \label{cha3-eq:poisson}
\end{equation}

In Fig.~\ref{cha3-fig:LSS} we report some representative ${\mathcal P}(S)$ for levels in the energy window around 
$\tilde{E}=0.26$ \footnote{We note that for level spacing distribution it is essential \cite{Sto99}
to distinguish levels from different symmetry classes. Here, the statistics is performed over 
the symmetric singlet states of the BHH. See also Ref.\cite{FFS89}.}.
One observes a qualitative change in the shape of ${\mathcal P}(S)$ from Poissonian-like associated
with very small and large $\lambda$ values to Wigner-like for intermediate values of $\lambda$.

In order to quantify the degree of level repulsion (and thus of chaoticity), various phenomenological 
formulas for ${\mathcal P}(S)$ have been suggested that interpolate between the two limiting cases 
(\ref{cha3-eq:GOE}, \ref{cha3-eq:poisson}) (see for example \cite{BR84,Brod73}). Here we use the 
so-called Brody distribution \cite{Brod73} given by the following expression
\begin{equation}
{\mathcal P}_{q}(S)=\alpha S^{q}\e{-\beta S^{1+q}}\szk\label{cha3-eq:brody}
\end{equation}
where $\alpha=(1+q)\beta$, $\beta=\Gamma^{1+q}[(2+q)/(1+q)]$ and $\Gamma$ is the Gamma function. 
The two parameters $\alpha,\,\beta$ are determined by the condition that the distribution is normalized
with a mean equal to one \cite{BFFMPW81}. The so-called Brody parameter $q$ is then obtained from 
direct fitting of ${\mathcal P}_{q}(S)$ to the numerically evaluated level spacing distribution. 
One readily verifies that for $q=0$, the distribution ${\mathcal P}_{q}(S)$ is Poissonian 
(\ref{cha3-eq:poisson}) while for $q=1$ it takes the form of (\ref{cha3-eq:GOE}). 

\begin{figure}[t]
\hfill{}\includegraphics[%
  width=0.9\columnwidth,
  keepaspectratio]{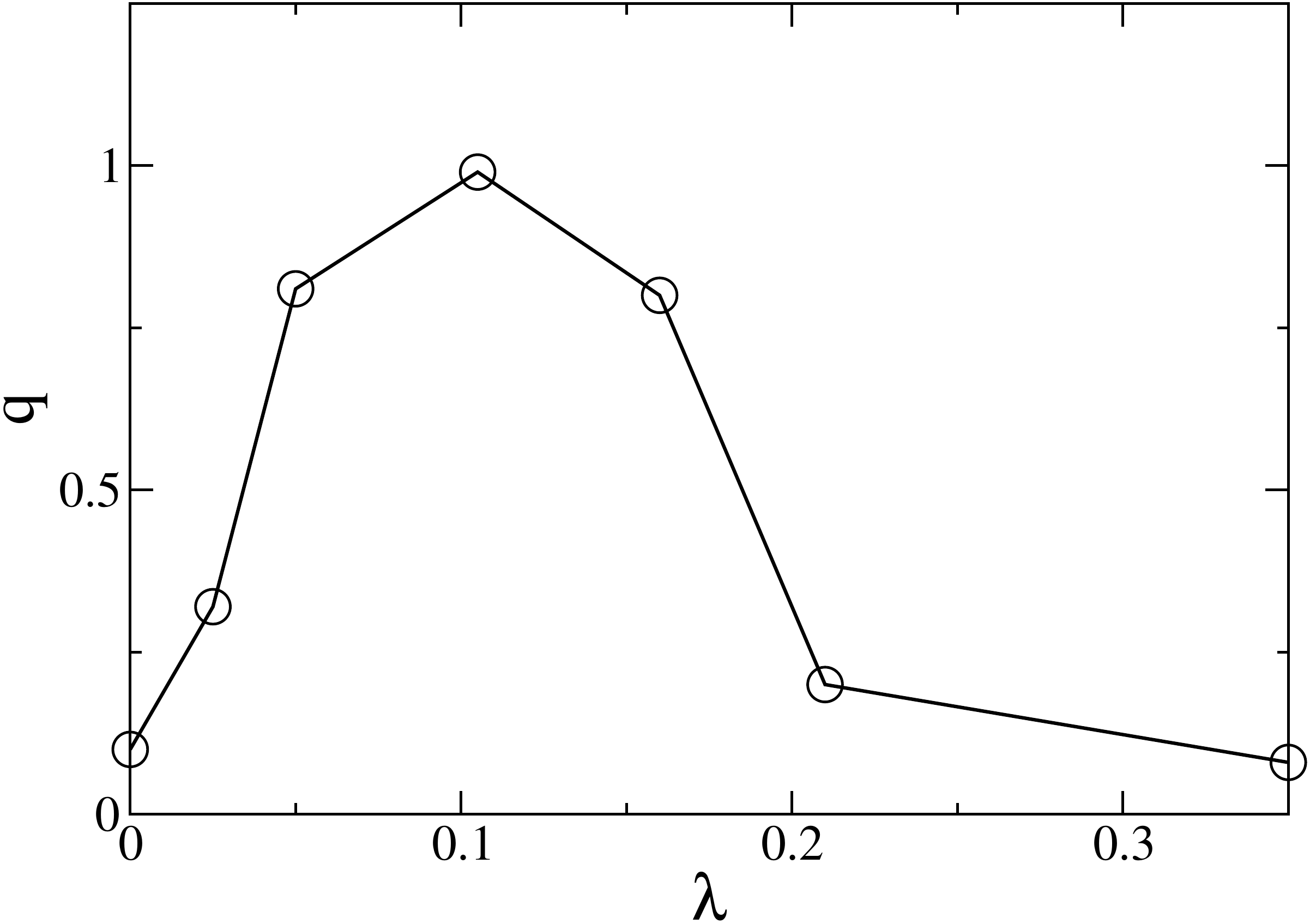}\hfill{}
\caption{\label{cha3-fig:LSS_brody}The Brody parameter $q$ for the BHH plotted
against the dimensionless ratio $\lambda$ which controls the underlying
classical dynamics. The values of $q$ are obtained from fits to $P_{q}(S)$
around $\tilde{E}=0.26$ as reported in Fig.~\ref{cha3-fig:LSS}. Error bars
are of the size of the circles. The System corresponds to $N=230$ bosons
and $\tilde{U}=280$. See text for details.}
\end{figure}

The fitted values of the Brody parameter $q$ for various $\lambda$'s are summarized in 
Fig.~\ref{cha3-fig:LSS_brody}. We see that for very small and very large $\lambda$ the Brody
parameter is small indicating classically regular motion while for intermediate values 
$0.04<\lambda <0.2$ we find $q\sim1$ corresponding to classically chaotic motion. 
This result is in perfect agreement with the predictions of the classical analysis.

%%%%%%    band profile   %%%%%%%%%%%%%%%
\subsection{The band profile\label{cha3-sec:bandprofile}}

The fingerprints of classically chaotic dynamics can be found also in the band-structure of the 
perturbation matrix ${\bf B}$. As we will show below the latter is related to the fluctuations 
of the classical motion. This is a major step towards a RMT modeling.

Consider a given ergodic trajectory $(I(\tilde{t}),\varphi(\tilde{t}))$ on the energy surface 
$\tilde{\mcal{H}}(I(0),\varphi(0);k_{0})=\tilde{E}$ (with $N=const.$). We can associate with it a stochastic-like 
variable 
\begin{equation}
\tilde{\mcal{F}}(\tilde{t})=-\frac{\partial\tilde{\mcal{H}}}{\partial k}(I(\tilde{t}),
\varphi(\tilde{t});k(\tilde{t}))\szk\label{cha3-eq:generalized-force}
\end{equation}
which can be seen as a generalized force. For the BHH (\ref{cha3-eq:BHH-hamilton_matrix})
this is simply given by the perturbation term i.e. 
\begin{equation}
\label{cha3-eq:generalized-force2}
\tilde{\mcal{F}}=\sum_{i\neq j}{\sqrt{I_{i}I_{j}}}\exp^{i(\varphi_{j}-\varphi_{i})}
\end{equation}
which corresponds to a \emph{momentum boost} since it changes the kinetic energy \cite{PBHJ07}. It 
may have a non-zero average, i.e.  a {}``conservative'' part, but below we are interested only in 
its fluctuations.

In order to characterize the fluctuations of $\tilde{\mcal{F}}(\tilde{t})$ we introduce the 
autocorrelation function 
\begin{equation}
C(\tilde{\tau})=\langle\tilde{\mcal{F}}(\tilde{t})\tilde{\mcal{F}}(\tilde{t}+\tilde{\tau})\rangle
-\langle\tilde{\mcal{F}}^{2}\rangle\szk
\label{cha3-eq:autocorrelation}
\end{equation}
where $\tilde{\tau}=\tilde{U}\tau$ is a rescaled time. The angular brackets denote an averaging which is either micro-canonical over some initial conditions 
$\left(I(0),\varphi(0)\right)$ or temporal due to the assumed ergodicity.

For generic chaotic systems (with smoothly varying potentials), the
fluctuations are characterized by a short correlation time $\tilde{\tau}_{\tbox{cl}}$,
after which the correlations are negligible. In generic circumstances
$\tilde{\tau}_{\tbox{cl}}$ is essentially the ergodic time. For our
system we have found $\tilde{\tau}_{\tbox{cl}}\sim2\pi$ [see Eq.~(\ref{cha3-eq:omega_cl})].

\begin{figure}
\hfill{}\includegraphics[%
  width=\columnwidth,
  keepaspectratio]{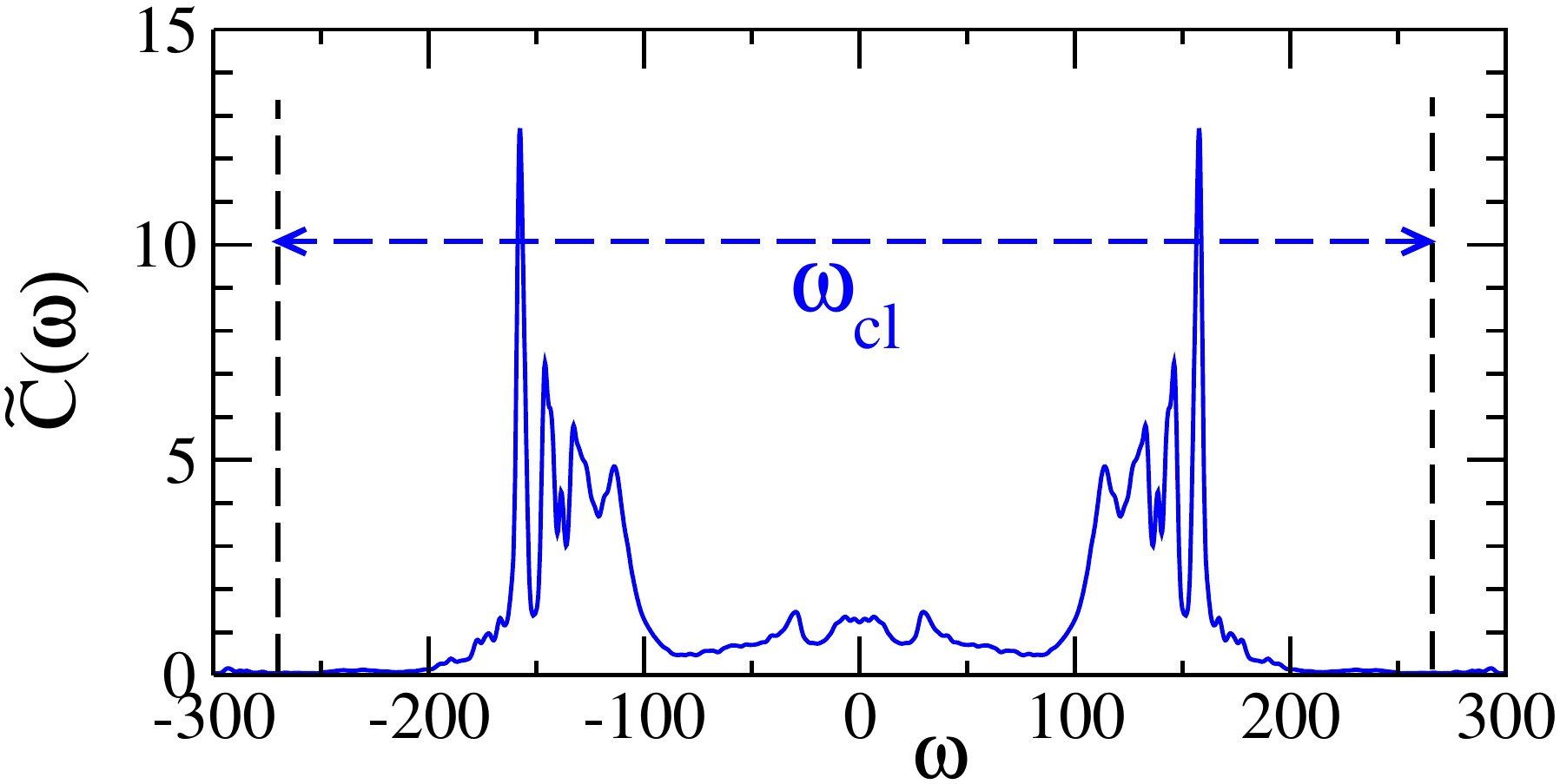}\hfill{}
\caption{\label{cha3-fig:clas-Cw}  (Color online)  The power-spectrum of the classical trimer
(\ref{eq:H-DNLS}) at energy $\tilde{E}=0.26$, $\tilde{U}=280$, and $\lambda_{0}=0.053$.
The classical cut-off frequency $\omega_{\tbox{cl}}=\tilde{\omega}_{\tbox{cl}}\tilde{U}\approx280$
is indicated by vertical dashed lines. }
\end{figure}

The power spectrum of the fluctuations $\tilde{C}(\tilde{\omega})$ is given by a Fourier 
transform: 
\begin{equation}
\tilde{C}(\tilde{\omega})=\int_{-\infty}^{\infty}C(\tilde{\tau})\, 
e^{i\tilde{\omega}\tilde{\tau}}\dd{\tilde{\tau}}\szk
\label{cha3-eq:power-spectrum}
\end{equation}
and for the case of the trimer (\ref{eq:H-DNLS}) is shown in Fig.~\ref{cha3-fig:clas-Cw}. We 
see that $\tilde{C}(\omega)$ has a (continuous) frequency support which is bounded by $\tilde{\omega}_{
\tbox{cl}}\approx1$ corresponding to $\omega_{\tbox{cl}}\approx280$ (indicated by dashed 
vertical lines in Fig.~\ref{cha3-fig:clas-Cw}). The cut-off frequency $\omega_{\tbox{cl}}$ is inverse 
proportional to the classical correlation time, i.e.
\begin{equation}
\tilde{\omega}_{\tbox{cl}}=\frac{2\pi}{\tilde{\tau}_{\tbox{cl}}}\szp
\label{cha3-eq:omega_cl}
\end{equation}
These characteristics of the power spectrum are \emph{universal}
for generic chaotic systems. Finally, we see that within the frequency
support the power spectrum $\tilde{C}(\tilde{\omega})$ is structured,
reflecting \emph{system-specific} properties of the underlying classical
dynamics. 

\begin{figure}
\hfill{}\includegraphics[%
  width=\columnwidth,
  keepaspectratio]{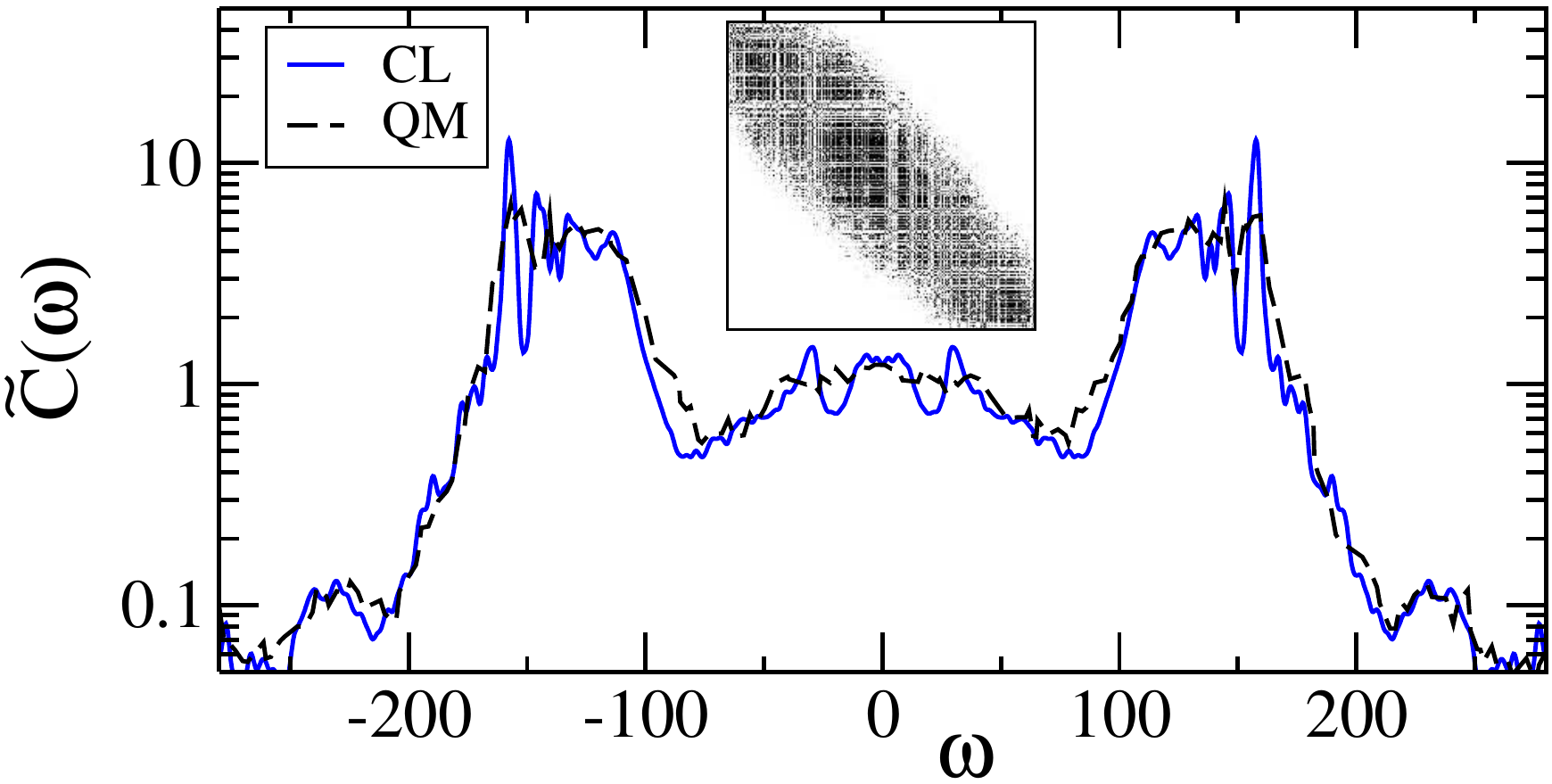}\hfill{}

\caption{\label{cha3-fig:bandprofile}  (Color online) The band profile $(2\pi\tilde{U}/N^{2}\Delta)\cdot|\mbf{B}_{nm}|^{2}$
versus $\omega=(E_{n}-E_{m})/\hbar$ is compared with the classical
power spectrum $\tilde{C}(\omega)$. The number of particles is $N=230$
and $\lambda_{0}=0.053$. \emph{Inset}: a snapshot of the perturbation
matrix $\mbf{B}_{nm}$.}
\end{figure}

The classical power spectrum $\tilde{C}(\tilde{\omega})$ is associated with the quantum 
mechanical perturbation matrix $\mbf{B}$ according to the following semiclassical relation 
\cite{FP86,PR93} 
\begin{equation}
\sigma_{nm}^2\equiv\langle|\mbf{B}_{nm}|^{2}\rangle=\frac{N^{2}\Delta}{\tilde{U}\,2\pi}\tilde{C}
\left(\omega=\frac{E_{n}-E_{m}}{\hbar}\right)\szp
\label{cha3-eq:bandprofile}
\end{equation}
Hence the matrix elements of the perturbation matrix $\mbf{B}$ are
extremely small outside a band of width 
\begin{equation}
b=\hbar\omega_{\tbox{cl}}/\Delta \approx \hbar\omega_{\tbox{cl}} N/{\tilde U}\szp
\label{cha3-eq:Delta_b_definition}
\end{equation}
In the inset of 
Fig.~\ref{cha3-fig:bandprofile} we show a snapshot of the perturbation matrix $|\mbf{B}_{nm}
|^{2}$ which clearly exhibits a band-structure. In the same figure we also display the scaled 
quantum band profile for $N=230$. The agreement with the classical power spectrum $\tilde{C}
(\omega)$ is excellent. We have checked that the relation (\ref{cha3-eq:bandprofile}) is very 
robust \cite{CK01,HCGK06,HKG06} and holds even for moderate number of bosons $N\approx50$. 
Combining Eqs.~(\ref{cha3-eq:Delta_exact}) and (\ref{cha3-eq:Delta_b_definition}) with $\tilde{
\omega}_{\tbox{cl}}\approx1$ (see above) and the definition of $b$ we find for the chaotic 
regime around $\tilde{E}=0.26$ that $b\sim0.6N$ which is confirmed by the numerics.

It is important to realize that upon quantization we end up with two distinct energy scales 
\cite{CK01,HCGK06,HKG06}. One is obviously the mean level spacing $\Delta\sim 1/N$  (see Eq.~
(\ref{cha3-eq:Delta_exact})) which is associated with the unperturbed Hamiltonian. The other 
energy scale is the bandwidth
\begin{equation}
\Delta_{b}=b\Delta\propto\tilde{U}\szk
\label{cha3-eq:bandwidth}
\end{equation}
which contains information about the power spectrum of the chaotic motion and is encoded in the
perturbation matrix ${\bf B}$. The latter energy scale is also known in the corresponding 
literature as the {}``non-universal'' 
energy scale \cite{Ber91}, or in the case of diffusive motion, as the Thouless energy \cite{Im97}.
One has to notice that deep in the semiclassical regime $N\rightarrow\infty$ these two energy scales 
differ enormously from one another. We shall see in the following sections that this scale
separation has dramatic consequences on the theory of wavepacket dynamics.

%%%%%%%%%%%%%%%%%  Distribution of coupling  %%%%%%%%%%%%%%%%%%%%%%%%%%%
\subsection{Distribution of matrix elements of the perturbation operator \label{cha3-sec:distribution_of_couplings}}

We further investigate the statistical properties of the matrix elements
$\mbf{B}_{nm}$ of the perturbation matrix by studying their distribution.
RMT assumes that upon appropriate {}``unfolding'' they must be distributed
in a Gaussian manner. The unfolding aims to remove system specific
properties and to reveal the underlying universality. It is carried
out by normalizing the matrix elements with the local standard deviation
$\sigma=\sqrt{\langle|{\textbf{B}}_{nm}|^{2}\rangle}$ related through
Eq.~(\ref{cha3-eq:bandprofile}) with the classical power spectrum
$\tilde{C}(\omega)$.

The existing literature is not conclusive about the distribution of
the normalized matrix elements $w=\mbf{B}_{nm}/\sigma$. Specifically,
Berry \cite{Berr77} and more recently Prosen \cite{PR93,Pros93b},
claimed that $\mathcal{P}(w)$ should be Gaussian. On the other hand,
Austin and Wilkinson \cite{AW92} have found that the Gaussian is
approached only in the limit of high quantum numbers while for small
numbers, i.e., low energies, a different distribution applies, namely
\begin{equation}
P_{\tbox{couplings}}(w)=\frac{\Gamma(\frac{N}{2})}{\sqrt{\pi N}
\Gamma(\frac{N-1}{2})}\left(1-\frac{w^{2}}{N}\right)^{(N-3)/2}\szp
\label{cha3-eq:wilkinson}
\end{equation}
 This is the distribution of the elements of an $N$-dimensional vector,
distributed randomly over the surface of an $N$-dimensional sphere
of radius $\sqrt{N}$. For $N\rightarrow\infty$ this distribution
approaches a Gaussian.

\begin{figure}
\hfill{}\includegraphics[%
  width=\columnwidth,
  keepaspectratio]{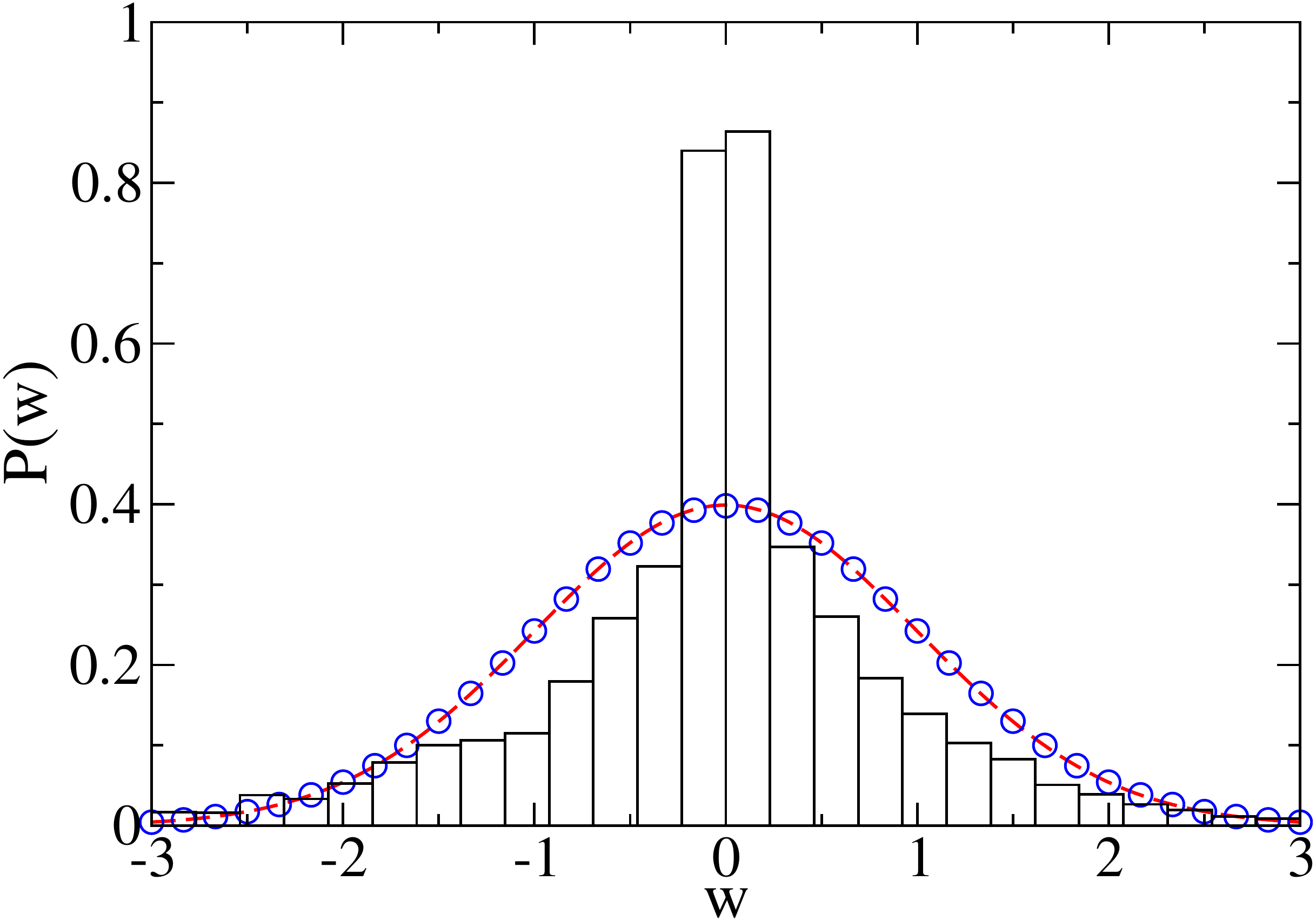}\hfill{}
\caption{\label{cha3-fig:BHHmatdist} (Color online) Distribution of rescaled matrix elements
$w$ around $\tilde{E}=0.26$ rescaled with the averaged band profile.
The dashed red line corresponds to the standard normal distribution
while the circles ($\circ$) correspond to a best fit from Eq.~(\ref{cha3-eq:wilkinson})
with a fitting parameter $N=342$. The system corresponds $N=230,\,\tilde{U}=280$.}
\end{figure}

In Fig.~\ref{cha3-fig:BHHmatdist} we report the distribution $\mathcal{P}(w)$
for the elements of the perturbation matrix $\mbf{B}$. The dashed line corresponds
to a Gaussian of unit variance while the circles are obtained by fitting
Eq.~(\ref{cha3-eq:wilkinson}) to the numerical data using $N$ as
a fitting parameter. Although we are deep in the semiclassical regime
(i.e. $N=230$), none of the above predictions describes in a satisfactory way the 
numerical data. We attribute these deviations to the existence of small 
stability islands in the phase space. Trajectories
started in those islands cannot reach the chaotic sea and vice versa.
Quantum mechanically, the consequence of this would be vanishing matrix
elements $\mbf{B}_{nm}$ which represent the classically forbidden
transitions.

%%%%  RMT  Modeling  %%%%%
\subsection{RMT modeling \label{cha3-sec:RMT} }

More than 50 years ago, E.~P.~Wigner \cite{Wign55,Wign57} proposed a simplified model to study 
the statistical properties of eigenvalues and eigenfunctions of complex systems. It is known 
as the \emph{Wigner banded random matrix} (WBRM) model. The corresponding Hamiltonian is given 
by Eq.~(\ref{cha3-eq:BHH-hamilton_matrix}) where $\mbf{B}$ is a \emph{banded} \textit{random} 
matrix \cite{FLW91,FGIM93,FCIC96}. This approach is attractive both analytically and numerically. 
Analytical calculations are greatly simplified by the assumption that the off-diagonal terms 
can be treated as independent random numbers. Also from a numerical point of view it is quite 
a tough task to calculate the true matrix elements of $\mbf{B}$. It requires a preliminary step 
where $\hat{H}_{0}$ is diagonalized. Due to memory limitations one ends up with quite small
matrices. For example, for the Bose-Hubbard Hamiltonian we were able to handle matrices of 
final size $\mcal{N}=30,000$ maximum. This should be contrasted with RMT simulations, where using 
self-expanding algorithm \cite{IKPT97,CIK00,HCGK06} we were able to handle system sizes up 
to $\mcal{N}=1,000,000$ along with significantly reduced CPU time. We would like to stress, 
however, that the underlying assumption of the WBRM, namely that the off-diagonal elements are 
\emph{uncorrelated} random numbers, has to be treated with extreme care. The applicability of 
this model is therefore a matter of conjecture which we will test in the following sections. 

In fact, the WBRM model involves an additional simplification. Namely, one assumes that the 
perturbation matrix $\mbf{B}$ has a \emph{rectangular} band profile of bandwidth $b$. A simple 
inspection of the band profile of our BHH model (see Fig.~\ref{cha3-fig:bandprofile}) shows 
that this is not the case. We eliminate this simplification by introducing a RMT model that is
even closer to the dynamical one. Specifically, we generate the matrix elements $B_{nm}$ from
a Gaussian distribution with a variance that is given by the classical power spectrum according
to Eq. (\ref{cha3-eq:bandprofile}). Thus the band-structure is kept intact. This procedure leads 
to a random model that exhibits only universal properties but lacks any classical limit. We 
will refer to it as the \emph{improved} random matrix theory model (IRMT).

%%%%%%%%%%%%%%%%  Pnm for trimeric BHH  %%%%%%%%%%%%%%%%%%%%%%%%%%
\section{Local Density of States and Quantum-Classical Correspondence\label{cha3-sec:P_nm_of_BHH}}
\begin{figure}
\hfill{}\includegraphics[%
  width=0.8\columnwidth,
  keepaspectratio]{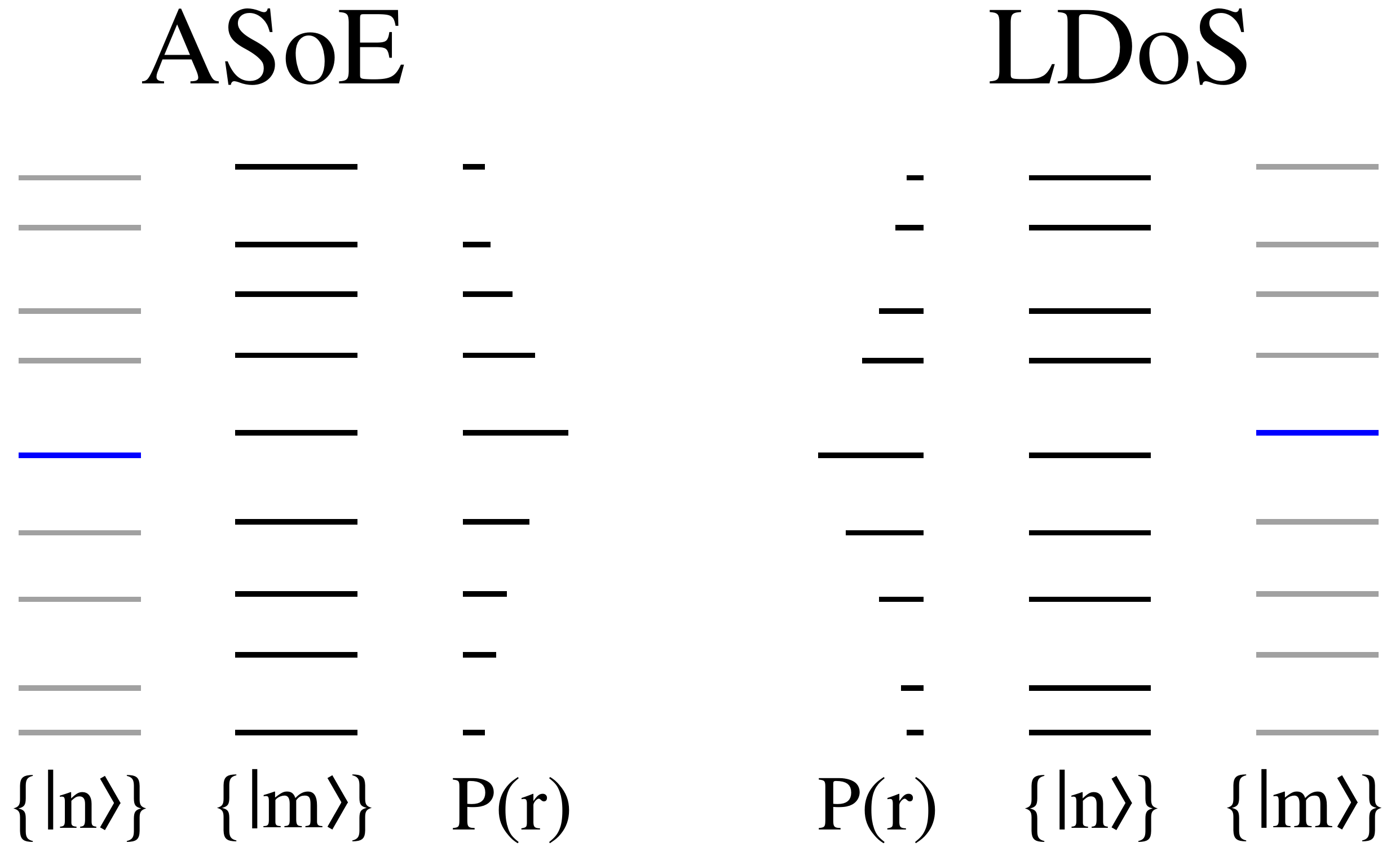}\hfill{}
\caption{\label{cha3-fig:P_r_scheme} (Color online) Schematic representation of the two notions
of the kernel $P(n|m)$. \emph{Left}: \emph{}Projection of one perturbed
eigenstate $|n(k_{0}+\delta k)\rangle$(blue level) on the basis $|m(k_{0})\rangle$
of the unperturbed Hamiltonian. Averaging over several $\tket{n'}$
states around energy $E_{n}$ yields the averaged shape of eigenfunctions
(ASoE). \emph{Right}: \emph{}Alternatively, if $P(n|m)$ is regarded
as a projection of one unperturbed eigenstate $|m\rangle$ (blue level)
on the basis $|n\rangle$ of the perturbed Hamiltonian and averaged
over several states around $E_{m}$ , it leads to the local density
of states (LDoS).}
\end{figure}

As we change the parameter $\delta k$ in the Hamiltonian (\ref{cha3-eq:BHH-hamilton_matrix}),
the instantaneous eigenstates $\{|n(k)\rangle\}$ undergo structural changes. Understanding
these changes is a crucial step towards the analysis of wavepacket dynamics \cite{CK01,HKG06}. 
This leads to the introduction of the {}``kernel'' 
\begin{eqnarray}
P(n|m)=|\langle n(k_{0}+\delta k)|m(k_{0})\rangle|^{2}\szk
\label{cha3-eq:kernel-QM}
\end{eqnarray}
which can be interpreted in two ways as we schematically depict in
Fig.~\ref{cha3-fig:P_r_scheme}. If regarded as a function of $m$,
$P(n|m)$ represents the overlap of a given perturbed eigenstate $|n(k_{0}+\delta k)\rangle$
with the eigenstates $|m(k_{0})\rangle$ of the unperturbed Hamiltonian.
The averaged distribution $P(r)$ is defined by ${r=n-m}$, and averaging
over several states with roughly the same energy $E_{n}$ yields the
averaged shape of eigenfunctions (ASoE). Alternatively, if regarded
as a function of $n$ and averaging over several states around a given
energy $E_{m}$, the kernel $P(r)$ represents up to some trivial
scaling and shifting the local density of states (LDoS): 
\begin{equation}
   P(E|m)=\sum_{n}|\langle n(k)|m(k_{0})\rangle|^{2}\delta(E-E_{n})\szp
\label{LDoS}
\end{equation}
Its line-shape is fundamental for the understanding of the associated
dynamics (see Sec.~\ref{cha4-sec:WP-dyn}), since its Fourier transform is the 
so-called {}``survival probability amplitude''.
In the following we will focus on the LDoS scenario.

\subsection{Parametric Evolution of the LDoS \label{cha3-sec:LDoS}}

\begin{figure}[!t]
\hfill{}\includegraphics[%
  width=\columnwidth,
  keepaspectratio]{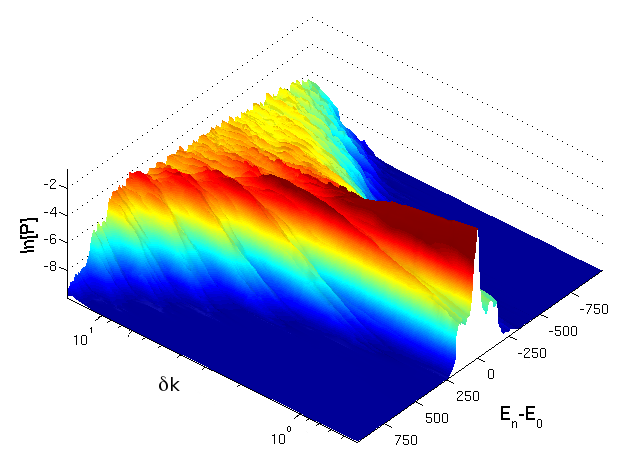}\hfill{}
\caption{\label{cha3-fig:Pnm_3D_plot} (Color online)  The kernel $P(n|m)$ of the BHH plotted
as a function of the perturbed energies $E_{n}$ (LDoS representation)
and for various perturbation strengths $\delta k>\delta k_{\textrm{qm}}$.
The averaged
shape of eigenfunctions is given by the same kernel $P(n|m)$ and
is obtained by just inverting the energy axis. Here, $N=70$, and
$\lambda_{0}=0.053$. }
\end{figure}

An overview of the parametric evolution of the averaged $P(n|m)$ is shown in 
Fig.~\ref{cha3-fig:Pnm_3D_plot} \cite{HKG06}. Beginning as a delta function for $\delta k=0$, 
the profile $P(n|m)$ starts to develop a non-perturbative core as $\delta k$ increases above 
some critical value $\delta k_{\tbox{qm}}$. For even stronger perturbations, $P(n|m)$ spills 
over the entire bandwidth $\Delta_{b}$. We will show that if $\delta k$ exceeds another critical 
value $\delta k_{\rm prt}$, the LDoS develops classical features. In the following 
we will identify the above parametric regimes and discuss the theory of $P(n|m)$ in
each one of them.

%%%  perturbative LDoS %%%
\subsubsection{The perturbative regimes}

\begin{figure}[!t]
\hfill{}\includegraphics[%
  width=1\columnwidth,
  keepaspectratio]{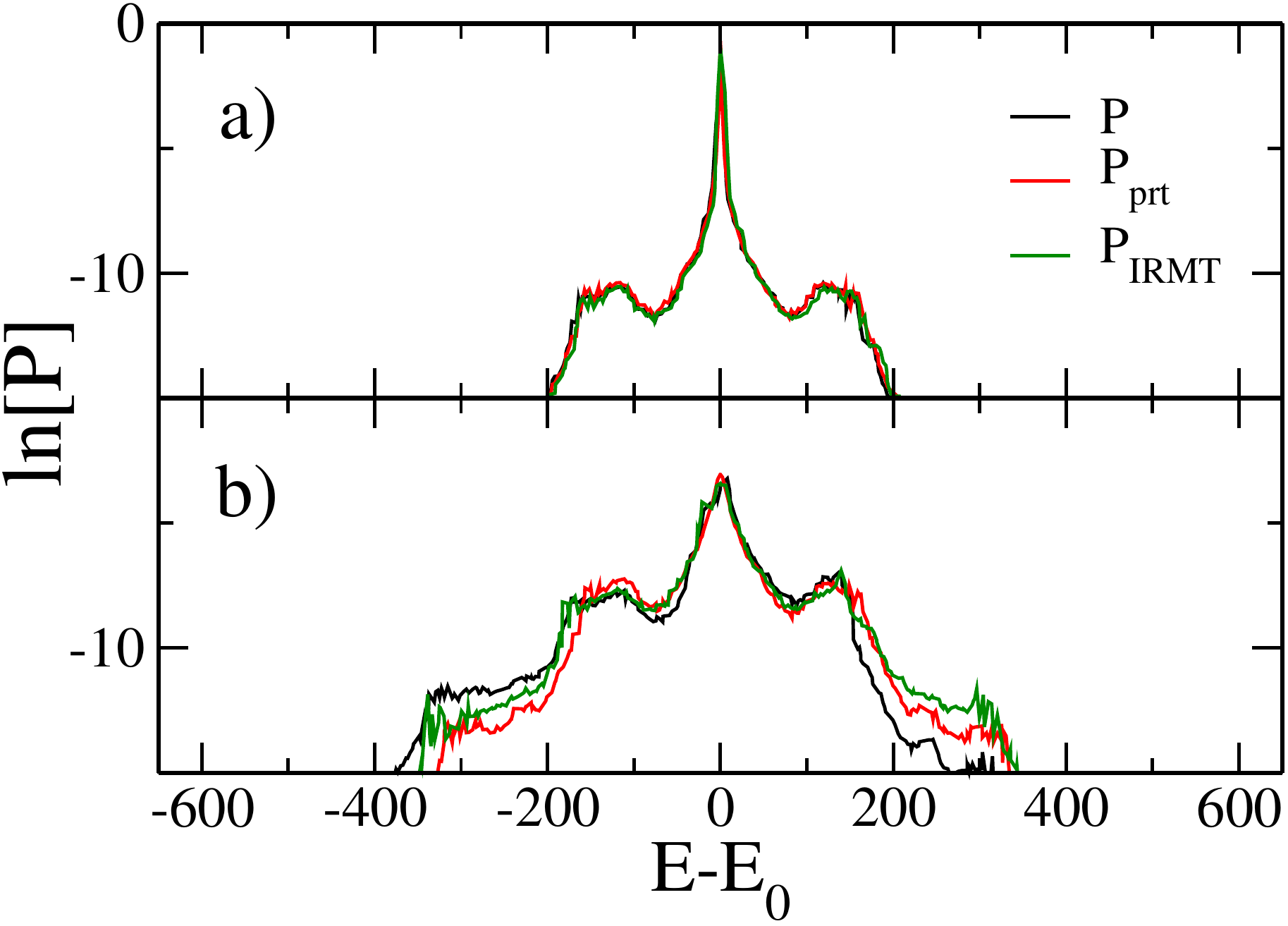}\hfill{}
\caption{\label{cha3-fig:Pnm-regimes_BHH}  (Color online) The quantal profile $P(n|m)$ as
a function of $E_{n}-E_{m}^{(0)}$ for the BHH model is compared with
$P_{\tbox{prt}}$ and with the corresponding $P_{\tbox{IRMT}}$ of
the IRMT model. The perturbation strength $\delta k$ is in (a) standard
perturbative regime $\delta k=0.05$ and (b) extended perturbative regime
$\delta k=0.3$. The system corresponds to $N=230$, $\tilde{U}=280$ and $k_{0}=15$.
Here $\delta k_{\textrm{qm}}=0.09$ and $\delta k_{\textrm{prt}}=1.02$. 
}
\end{figure}

We start with the discussion of the perturbative regimes. We distinguish between two cases:

{\it The Standard Perturbative Regime}:
The simplest case is obviously the first order perturbation theory (FOPT) \emph{}regime 
where, for $P(n|m)$, we can use the standard textbook approximation $P_{\tbox{FOPT}}(n|m)\approx1$ for $n=m$, 
while 
\begin{equation}
P_{\tbox{FOPT}}(n|m)=\frac{\delta k^{2}\,\,|\mbf{B}_{mn}|^{2}}{(E_{n}-E_{m})^{2}}
\label{cha3-eq:P_FOPT}
\end{equation}
for $n\ne m$. The border $\delta k_{\tbox{qm}}$ for which Eq. (\ref{cha3-eq:P_FOPT}) describes the
LDoS kernel, can be found by the requirement that only nearest-neighbor levels are mixed by the 
perturbation. We get 
\begin{equation}
\delta k_{\tbox{qm}}=\Delta/\sigma \propto\frac{\tilde{U}}{N^{3/2}}\szk
\label{cha3-eq:delta k_qm}
\end{equation}
where for the rhs. of Eq.~(\ref{cha3-eq:delta k_qm}) we have used the scaling relations for 
$\Delta$ and $\sigma$ (see Eqs.~(\ref{cha3-eq:Delta_exact}) and (\ref{cha3-eq:bandprofile})). In 
Fig.~\ref{cha3-fig:Pnm-regimes_BHH}a we report our numerical results for the BHH, together with the 
perturbative profile $P_{\tbox{FOPT}}(n|m)$ obtained from Eq.~(\ref{cha3-eq:P_FOPT}) and the outcome
of the IRMT modeling. The FOPT Eq.~(\ref{cha3-eq:P_FOPT}) has as an input the classical power spectrum 
${\tilde C}(\omega)$ which via Eq. (\ref{cha3-eq:bandprofile}) can be used in order to evaluate the 
band profile ${\bf B}_{nm}$. All three curves fall on top of one another.

{\it Extended Perturbative Regime}: If $\delta k>\delta k_{\tbox{qm}}$ but not too large then we 
expect that several levels are mixed non-perturbatively. This leads to a distinction between a {}``core'' 
of width $\Gamma$ which contains most of the probability and a tail region which is still described by 
FOPT. This non-trivial observation can be justified using perturbation theory to infinite order. It 
turns out that the non-perturbative mixing on the small scale $\Gamma$ of the core does not affect 
the long-range transitions \cite{CH00,CK01} that dictate the tails. Therefore we can argue that a 
reasonable approximation is \cite{CK01}
\begin{equation}
P_{\tbox{prt}}(n|m)=\frac{\delta k^{2}\,\,|\mbf{B}_{mn}|^{2}}{(E_{n}-E_{m})^{2}+\Gamma^{2}}.
\label{cha3-eq:P_FGR}
\end{equation}
Our numerical data, reported in Fig.\ref{cha3-fig:Pnm-regimes_BHH}b, indicate again an excellent
agreement with the theoretical prediction (\ref{cha3-eq:P_FGR}). At the same time, we observe that also the
proposed IRMT describes quite nicely the actual profile $P(r)$. Note that the resulting 
line-shape is strikingly different from a Wigner Lorentzian (as predicted by the traditional RMT 
modeling) and is rather governed by the semiclassical structures of the band profile 
$|{\bf B}_{nm}|^2$. Instead, a Wigner Lorentzian would be obtained if the band profile of the 
perturbation matrix were flat. 

\begin{figure}
\hfill{}\includegraphics[%
  width=0.9\columnwidth,
  keepaspectratio]{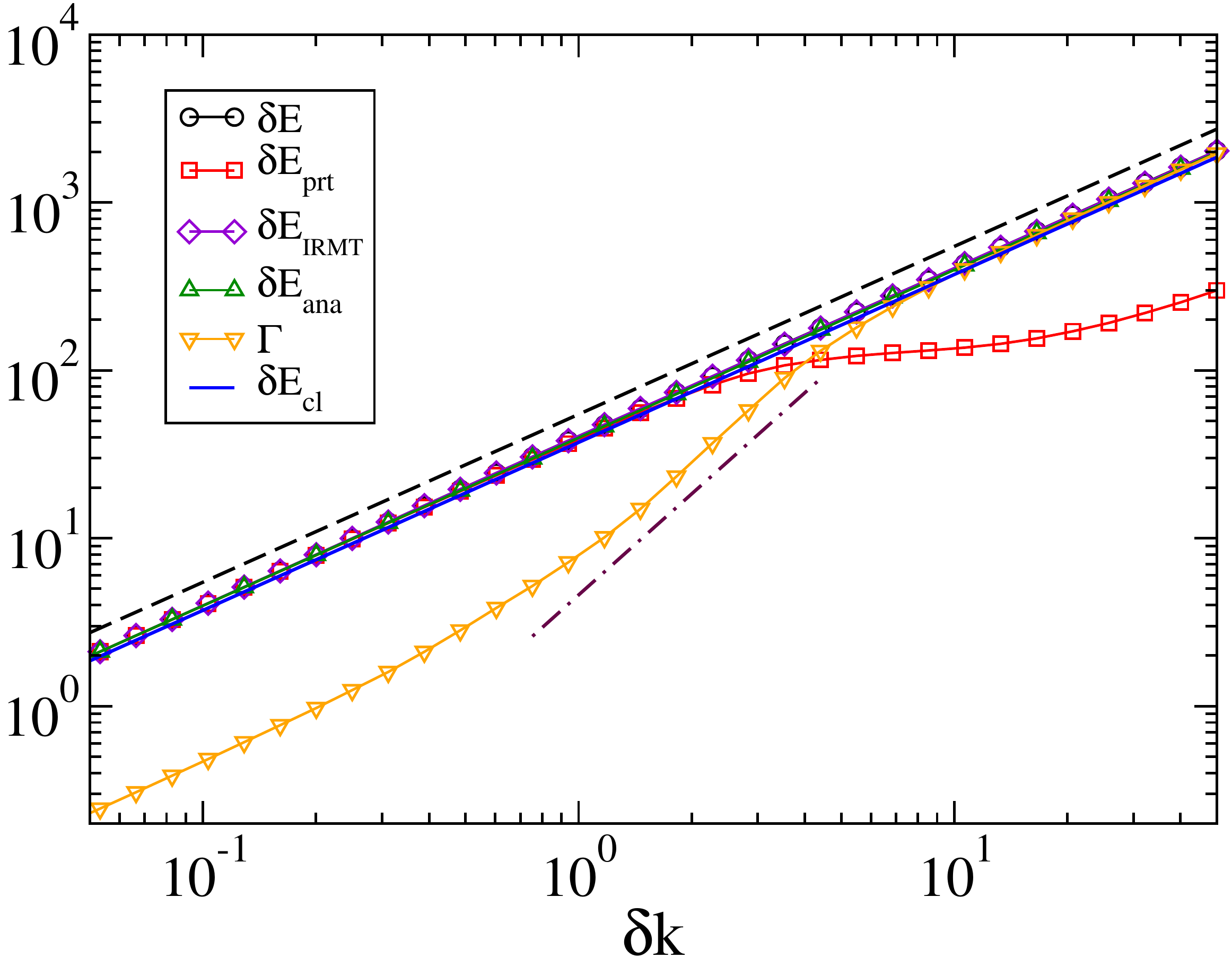}\hfill{}
\caption{\label{cha3-fig:dE_qm_cl_IPR_etc} (Color online)  Various measures of the spreading
profile for the BHH and the IRMT model: the quantal spreading $\delta E$
(black line), the quantal spreading $\delta E_{\tbox{prt}}$ of the
perturbative profile given by Eq.~(\ref{cha3-eq:P_FGR}) (red line), the spreading
$\delta E_{\rm IRMT}$ obtained from the IRMT modeling,
the analytical spreading $\delta E_{\tbox{ana}}$ obtained from (\ref{cha3-eq:dE_qm_ana})
(green line), the core-width $\Gamma$ (orange line), and the classical
spreading $\delta E_{\tbox{cl}}$ (blue line). The dashed line has
slope one, while the dash-dotted line has slope two and are drawn
to guide the eye. The systems correspond to $N=70$ bosons, $k_{0}=15$
and $\tilde{U}=280.$ See text for details.
}
\end{figure}

The core-width $\Gamma$ is evaluated by imposing normalization on $P_{\tbox{prt}}(n|m)$ \cite{HKG06}. 
Our numerically evaluated $\Gamma$ is reported in Fig. \ref{cha3-fig:dE_qm_cl_IPR_etc}. We see that 
for very small $\delta k$ we get that $\Gamma\ll \Delta$. In this case, the expression (\ref{cha3-eq:P_FGR}) 
collapses to the FOPT expression (\ref{cha3-eq:P_FOPT}). In fact, the inequality $\Gamma\leq \Delta$ 
can be used in order to estimate the limit $\delta k_{\tbox{qm}}$ of the validity of FOPT. As soon as we enter 
the extended perturbative regime, we find (see Fig. \ref{cha3-fig:dE_qm_cl_IPR_etc}) that $\Gamma$ 
grows as
\begin{equation}
\Gamma\propto\left(\delta k\frac{\sigma}{\Delta}\right)^2\times\Delta\quad.
\label{cha3-eq:gamma_bounds_BHH}
\end{equation}
The core-width $\Gamma$ (and thus Eq.~(\ref{cha3-eq:P_FGR}) for the LDoS) is meaningful only as 
long as we have $\Gamma<\Delta_{b}$, i.e. as long as we can distinguish a core-tail structure. 
This condition allows us to evaluate the perturbative border $\delta k_{\tbox{prt}}$: 
\begin{equation}
\delta k_{\tbox{prt}}\propto\frac{\tilde{U}}{N}\szp
\label{cha3-eq:dk_prt_BHH}
\end{equation}

\begin{figure}
\hfill{}\includegraphics[%
  width=\columnwidth,
  keepaspectratio]{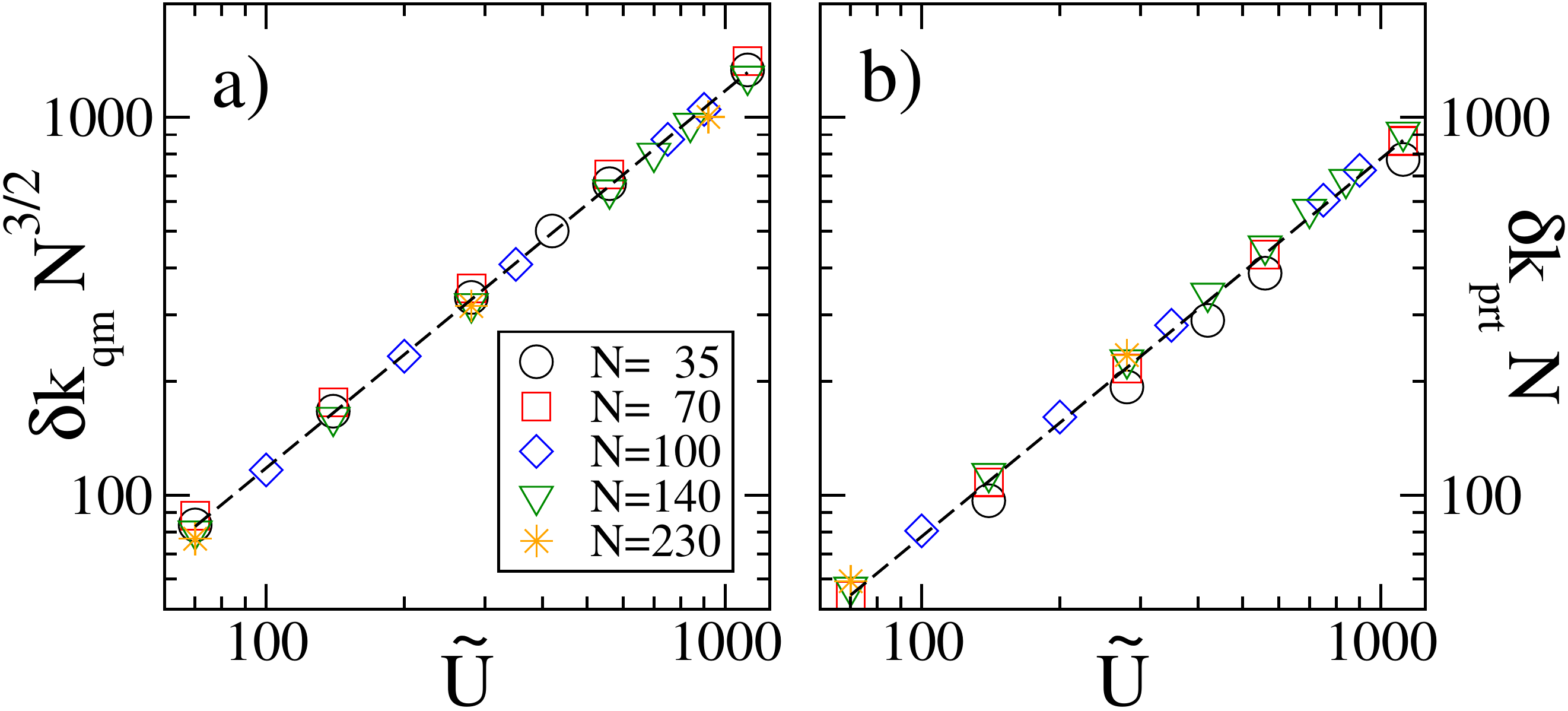}\hfill{}
\caption{\label{cha3-fig:prt_scaling}  (Color online) The parameters (a) $\delta k_{\textrm{qm}}$
and (b) $\delta k_{\textrm{prt}}$ for various $\tilde{U},N,$ and
for $\lambda_{0}=0.053$. A nice scaling in accordance with Eqs.(\ref{cha3-eq:delta k_qm})
and (\ref{cha3-eq:dk_prt_BHH}) is observed. }
\end{figure}

In our numerical analysis we have defined $\delta k_{\tbox{qm}}$ as the perturbation strength for 
which $50\%$ of the probability remains at the original site but we have checked that the condition
$\Gamma=\Delta$ gives the same result. For determining $\delta k_{\tbox{prt}}$ we use the following 
numerical procedure: We calculate the spreading $\delta E=\sqrt{\sum_{n}P(n|m) (E_{m}^{(0)}-E_{n}
)^{2}}$ of $P(r)$. Next we calculate $\delta E_{\tbox{prt}}(\delta k)$, using Eq.(\ref{cha3-eq:P_FGR})). 
This quantity always saturates for large $\delta k$ because of having a finite bandwidth. We compare 
it to the exact $\delta E(\delta k)$ and define $\delta k_{\tbox{prt}}$, for instance, as the $80\%$ 
departure point. In Fig.~\ref{cha3-fig:prt_scaling}, we present our numerical data for $\delta k_{
\rm qm}$ and $\delta k_{\rm prt}$ by making use of the scaling relations (\ref{cha3-eq:delta k_qm}) and
(\ref{cha3-eq:dk_prt_BHH}). A nice overlap is evident, confirming the validity of the above
expressions.

%%%%%%%%%%%%%%%%% non-perturbative LDoS  %%%%%%%%%%%%%%%%%%%%%%%%%%%%%%%%%%%%%%%%%%%%%%
\subsubsection{The non-perturbative regime \label{cha3-sec:QCC_of_BHH}}

\begin{figure}
\hfill{}\includegraphics[%
  width=\columnwidth,
  keepaspectratio]{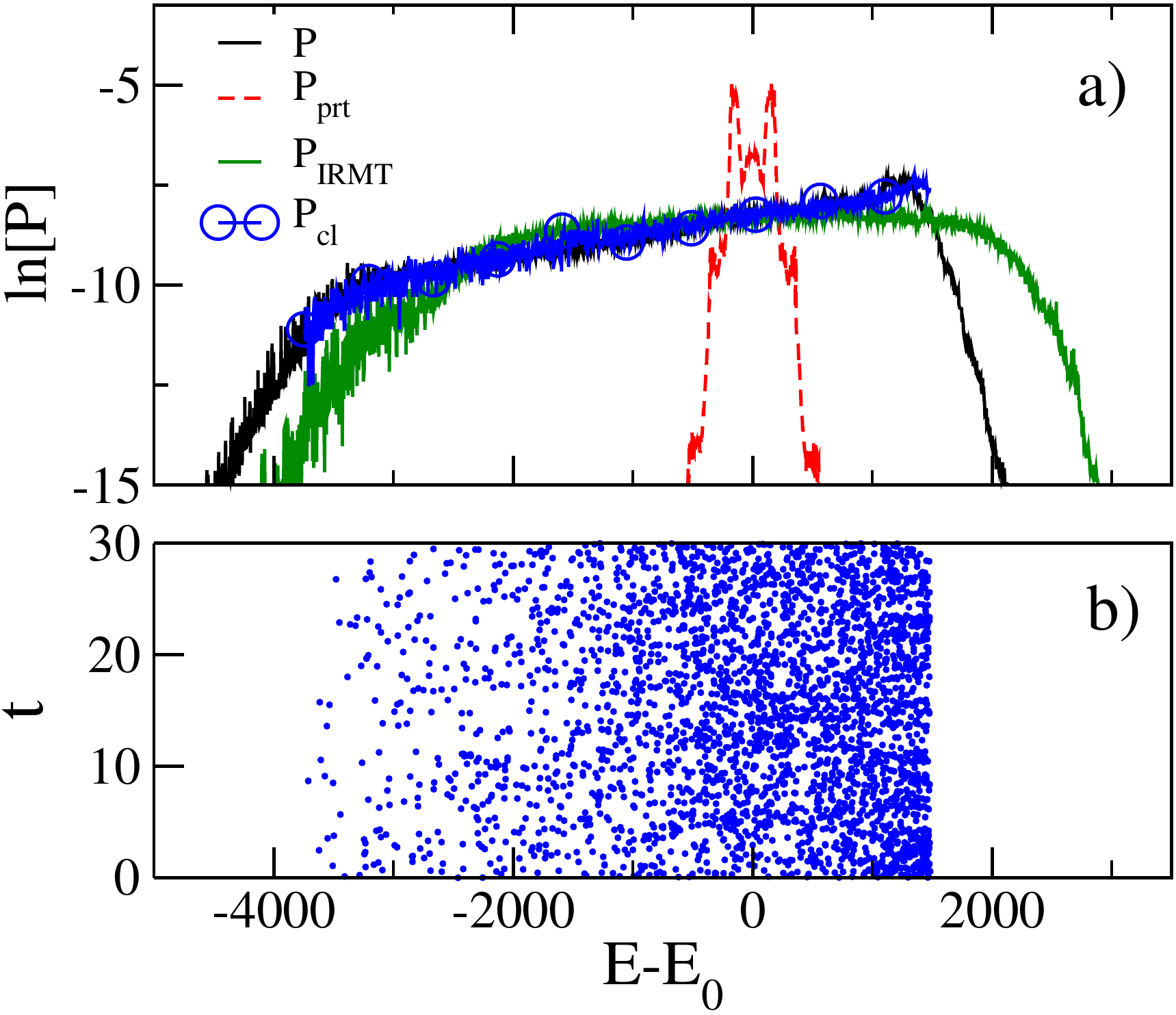}\hfill{}
\caption{\label{cha3-fig:P_cl_histogram} (Color online) {\em Upper panel}: The kernel $P(n|m)$ (LDoS representation)
in the non-perturbative regime $\delta k=10$ for $N=230$ and $\lambda=0.053$. The results of the
BHH model (solid black line) are compared with $P_{\rm prt}$ (red dashed line), $P_{\tbox{IRMT}}$ of the IRMT
model (solid green line), and the classical profile $P_{\tbox{cl}}$ (blue line with $\circ$).
{\em Lower panel}: A time series $E(t)$ which leads to the classical profile $P_{\tbox{cl}}(E)$ (see text
for details). }
\end{figure}

For $\delta k>\delta k_{\rm prt}$ the core spills over the bandwidth and therefore perturbation 
theory, even to infinite order, is inapplicable for evaluating $P(n|m)$. In this regime, we 
observe that also the IRMT fails to reproduce the actual line-shape of $P(n|m)$ as can be seen 
from Fig.~\ref{cha3-fig:P_cl_histogram}a. In fact, RMT modeling leads to a semicircle 
\begin{equation}
\label{Wsmc}
P(n|m) = 1/(2\pi\Delta) \sqrt{4-((E_n-E_m)/\Delta)^2}
\end{equation}
as predicted by Wigner \cite{Wign55}. 

What is the physics behind the LDoS line-shape in the non-perturbative regime? Due to the strong 
perturbations many levels are mixed and hence the quantum nature becomes {}``blurred''. Then, we 
can approximate the spreading profile by the semiclassical expression \cite{CH00,CK01,MKC05}
\begin{equation}
P_{\tbox{sc}}(n|m)\,\,=\,\,\int{\frac{dI\, d\varphi}{(2\pi\hbar)^{d}}}\rho_{n}(I,\varphi)
\rho_{m}(I,\varphi)\szk
\label{cha3-eq:P_semiclassical}
\end{equation}
where $\rho_{m}(I,\varphi)$ and $\rho_{n}(I,\varphi)$ are the Wigner functions that correspond to 
the eigenstates $|m(k_{0})\rangle$ and $|n(k)\rangle$ respectively. In the strict classical limit 
$\rho$ can be approximated by the corresponding micro-canonical distribution $\rho\propto\delta(
E{-}{\mathcal{H}}(\{I_i\},\{\varphi_i\}))$ determined by the energy surface $E$. The latter can be 
evaluated by projecting the dynamics generated by ${\cal H}_0(\{I_i\},\{\varphi_i\})=E_0$ onto the 
Hamiltonian ${\cal H}(\{I_i\}, \{\varphi_i\})=E(t)$.

In Fig.~\ref{cha3-fig:P_cl_histogram}b we plot the resulting $E(t)=\mcal{H}(I(t),\varphi(t))$ as a 
function of time for the Hamiltonian (\ref{eq:H-DNLS}). The classical distribution $P_{\tbox{cl}}(n|m)$ 
is constructed (Fig.~\ref{cha3-fig:P_cl_histogram}a) from $E(t)$, by averaging over a sufficiently 
long time. The good agreement with the quantum profile $P(n|m)$ is a manifestation of the detailed 
quantum-classical correspondence which affects the whole LDoS profile in the non-perturbative regime. 

Coming back to the failure of the IRMT approach, we are now able to understand it formally from the
scaling relation (\ref{cha3-eq:dk_prt_BHH}) of the perturbative border $\delta k_{\tbox{prt}}\sim
\tilde{U}/N$. Specifically, we observe that the non-perturbative limit can be approached either by 
increasing the perturbation strength $\delta k$ or, alternatively, by keeping $\delta k$ constant
and increasing $N$. As we have seen before increasing $N$ means to approach the classical limit 
(keeping $\tilde{U}=const.$). On the other hand, it is clear that the IRMT model lacks a classical 
limit! Therefore, we cannot expect it to yield a correct description of $P(n|m)$ in that regime. Instead, 
for $\delta k>\delta k_{\tbox{prt}}$ the LDoS is completely dictated by semiclassical considerations as
can be seen from Fig.~\ref{cha3-fig:P_cl_histogram}a.

%%%%%%%%%%%%%%%  Second Moment and  QCC  %%%%%%%%%%%%%%%%%%%%%%%%%%%%%%%
\subsection{Restricted vs. Detailed Quantum-Classical Correspondence \label{cha3-sec:QCC}}

It is important to distinguish between detailed and restricted quantum-classical correspondence
(QCC) \cite{Cohe00,CK00}. The two types of QCC are defined as follows: (a) detailed QCC means
$P(r) \approx P_{\tbox{cl}}(r)$ while (b) restricted QCC means $\delta E_{\tbox{qm}} \approx \delta E_{\tbox{cl}}$. 

Obviously restricted QCC is a trivial consequence of detailed QCC, but the converse is not true. 
It turns out that restricted QCC is much more robust than detailed QCC. In Fig.~\ref{cha3-fig:dE_qm_cl_IPR_etc}
we see that the dispersion $\delta E_{\tbox{qm}}$ of either $P(r)$ 
or $P_{\tbox{IRMT}}(r)$ is almost indistinguishable from $\delta E_{\tbox{cl}}$. In fact, this 
agreement of the second moment $\delta E$ persists also for the case of the perturbative profile 
(\ref{cha3-eq:P_FGR}). This is quite remarkable because the corresponding LDoS profiles (quantal,
perturbative, IRMT and classical) can become very different!

The possibility of having restricted QCC was pointed out in \cite{CK01,MKC05} in the frame of quantum 
systems with chaotic classical limit. A simple proof presented in Ref.~\cite{CK01} indicated that
the variance of $P(r)$ is determined by the first two moments of the Hamiltonian in the unperturbed 
basis i.e.
\begin{eqnarray}
  \delta E^{2} & = & \langle m|\hat{H}^{2}|m\rangle-\langle m|\hat{H}|m\rangle^{2}\nonumber\\
  & = &\delta k^{2}\left[\langle m|\hat{B}^{2}|m\rangle-\langle m|\hat{B}|m\rangle^{2}\right]\nonumber \\
  & = & \delta k^{2}\left[\sum_{n}|\mbf{B}_{nm}|^{2}-|\mbf{B}_{mm}|^{2}\right]\szp
  \label{cha3-eq:dE_qm_ana}
\end{eqnarray}
Having a  $\delta E_{\tbox{qm}}$ that is determined only by the band profile, is the reason for restricted 
QCC, and is also the reason why restricted QCC is not sensitive to the RMT assumption.

%%%%%%%%%%%%   WAVEPACKET DYNAMICS   %%%%%%%%%%%%%%%%%%%%%%%%%%%%%%%%%%%%%
\section{Wavepacket dynamics \label{cha4-sec:WP-dyn}}

We now turn to the time-dependent scenario of the wavepacket dynamics which is related to the
response  of a system to a rectangular pulse. Its physical realization in the framework of the BHH
has been described in Section \ref{sec:object}. 

In the next subsections we will discuss the time-evolving energy profile in each of the three $\delta k$-regimes
which we have identified in the frame of the LDoS study. We start our analysis with the classical dynamics 
(Subsection \ref{cha4-sec:WPD-clas}) and then turn to the evolution of the quantum profile $P_{t}(r)$ 
(Subsection \ref{cha4-sec:WPD-regimes}). In the same subsection we will present an analysis
of the IRMT and semiclassical modeling and identify both their weakness and regimes of validity. 

%%%%%%%%%%%%%%%%%%%%%%%%%%  classical dynamics  %%%%%%%%%%%%%%%%%%%%%%%%%%%%%%%%
\subsection{Classical Dynamics \label{cha4-sec:WPD-clas}}

The classical picture is quite clear: The initial preparation is assumed to be a micro-canonical 
distribution that is supported by the energy surface ${\mathcal{H}}_{0}(I,\varphi)= E(0)=E_{n_{0}}$ 
where the Hamiltonian is given by Eq.~(\ref{eq:H-DNLS}). Taking ${ {\cal H}}(\lambda)$ to be 
a generator for the classical dynamics, the phase-space distribution spreads away from the initial 
surface for $t>0$. {}``Points'' of the evolving distribution move upon the energy surfaces of 
${\mathcal{H}}(I,\varphi)$. Thus, the energy $E(t)={\mathcal{H}}_{0}(I(t),\varphi(t))$ of the
evolving distribution spreads with time. We are interested in the distribution of ${ E}(t)$ of 
the evolving {}``points". 

A quantitative description of the classical spreading is easily obtained from Hamilton's equations: 
\begin{equation}
{\dd{{ E}}({ t})\over \dd{{ t}}}= [{ {\cal H}},{ {\cal H}}]_{\tbox{PB}} 
+ {{\partial  {\mathcal H}} \over {\partial { t}}} = -\delta k \dot{f}({ t}) \mathcal{F}({t})
\end{equation}
where $[ \cdot ]_{\rm PB}$ indicates the Poisson Brackets and $f(t)$ is a rectangular pulse i.e. 
$f(t')=1$ for $0<t'<t$. Integrating the previous expression and then taking a micro-canonical 
average over initial conditions we get for the energy spreading the classical linear response 
theory (LRT) expression 
\begin{equation} 
\delta { E}_{\tbox{cl}}({t}) = \delta k \times \sqrt{2[C(0)-C({t})]}
\approx\left\{
\begin{array}{lcr}
\delta { E}_{\rm cl}{{ t}\over { \tau}_{\rm cl}}\, ; & { t}<{ \tau}_{\rm cl}\\
\delta { E}_{\rm cl}\, ;  & { t}>{ \tau}_{\rm cl}
\end{array}\right.
\label{eq:LRT-dE-wpk2}.
\end{equation}
In the last step, we have expanded the correlation function for ${ t}\ll { \tau}_{\rm cl}$ as 
$C({ t})\approx C(0)-{1\over2}C''(0){ t}^2$. For ${ t}\gg{ \tau}_{\rm cl}$, due to ergodicity,
a {}``steady-state distribution" appears, where the evolving 
{}``points" occupy an {}``energy shell" in phase-space. The thickness of this energy 
shell equals $\delta { E}_{\tbox{cl}}$. Thus, the classical dynamics is fully characterized 
by the two classical parameters ${ \tau}_{\tbox{cl}}$ and $\delta { E}_{\tbox{cl}}$.

\begin{figure}
\hfill{}\includegraphics[%
  width=0.9\columnwidth,
  keepaspectratio]{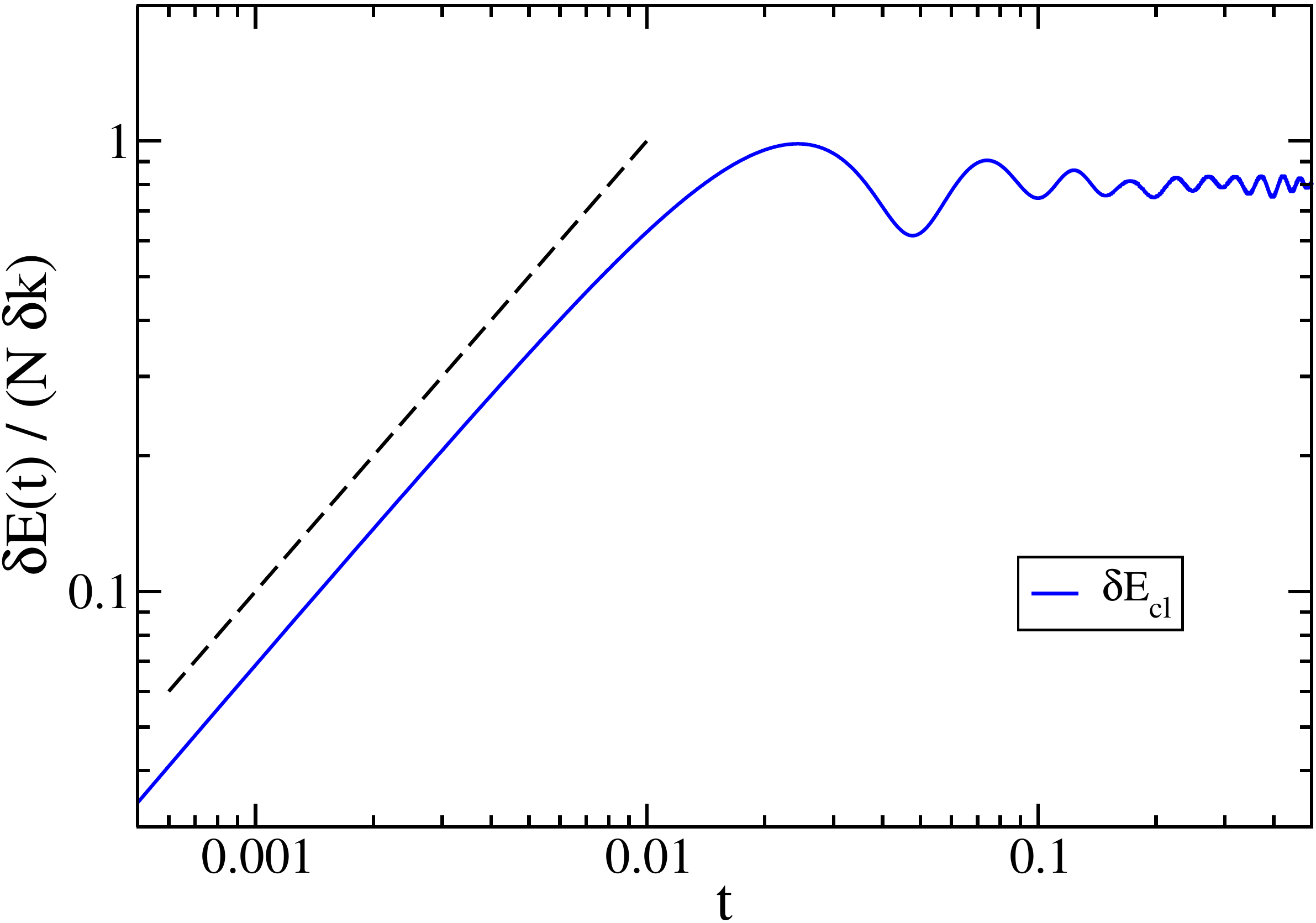}\hfill{}
\caption{\label{cha4-fig:dE_cl} (Color online)  The classical energy spreading $\delta E_{\tbox{cl}}(t)$
for the BHH (normalized with respect to the perturbation strength $\delta k$ and the boson 
number $N$) is plotted as a function of time. The dashed line has slope one and is drawn to 
guide the eye.}
\end{figure}

Figure \ref{cha4-fig:dE_cl} shows the scaled classical energy spreading $\delta E_{\tbox{cl}}(t)/
(N\,\delta k)$ for the BHH. The heavy dashed line has slope one and is drawn to guide the eye. 
In agreement with Eq.~(\ref{eq:LRT-dE-wpk2}) we see that $\delta E_{\tbox{cl}}(t)$ is first 
ballistic and then saturates at $\tau_{\tbox{cl}}\approx 2\pi/{\tilde U} = 0.02$. 

One can also calculate the entire classical evolving profile $P_{\rm cl}(t)$. Using a phase-space
approach similarly to the LDoS case in Subsection~\ref{cha3-sec:QCC_of_BHH} we propagate up to time 
$t$ under the Hamiltonian $\mcal{H}$, a large set of trajectories $\{ E\}_{t=0}$ that originally are 
supported by the energy surface ${\mathcal{H}}_{0}(I,\varphi)=E(t=0)=E_{n_{0}}$. Projecting them
back onto $\mcal{H}_{0}$ yields a set of energies $\{ E\}_{t=t}$ whose
distribution \footnote{Technically, this requires calculating the histogram with a bin-size
given by the mean level spacing $\Delta$.} constitutes the spreading profile $P_{\tbox{cl}}(t)$ at
time $t$. We will discuss  $P_{\tbox{cl}}(t)$ in Subsection \ref{cha4-sec:WPD-detail}.

%%%%%%%%%%%%%%%%%%%%%  quantum dynamics  %%%%%%%%%%%%%%%%%%%%%%%%%%%%
\subsection{Quantum Dynamics \label{cha4-sec:WPD-regimes}}

Now we would like to explore the various dynamical scenarios that are generated by the 
Schr\"odinger equation for $a_n(t)=\langle n|\psi(t) \rangle$. Namely, we want to solve
\begin{eqnarray} 
\label{eq3}
\frac{da_n}{dt} \ = \
-\frac{i}{\hbar} E_n \ a_n
\ -\frac{i}{\hbar}\sum_{m} \mbf{B}_{nm} \ a_{m}\szk
\end{eqnarray}
starting with an initial preparation $a_n=\delta_{nm}$ at $t{=}0$, i.e. an eigenstate of the
unperturbed system. We describe the energy 
spreading profile for $t>0$ by the transition probability kernel $P_t(n|m)=\langle |a_n(t)|^2 
\rangle$. The angular brackets stand for averaging over initial states ($m$) belonging to 
the energy interval ${0.25} \le {\tilde E_m}\le { 0.27}$. We 
characterize the evolving distribution using the various measures introduced in subsection 
\ref{sec:measures}. If the evolution is classical-like then --according to the classical analysis 
presented previously--  $P_t(n|m)$ will be characterized by a single energy scale 
$\delta E(t)$, meaning that any other measure like $\delta E_{\rm core}(t)$ reduces (up 
to a numerical factor) to $\delta E(t)$. We will use this criterion in the following in order 
to identify for which $\delta k$-regimes the evolution is classical-like and for which ones it
develops quantum features.

\begin{figure}[t!]
\hfill{}\includegraphics[%
  width=\columnwidth,
  keepaspectratio]{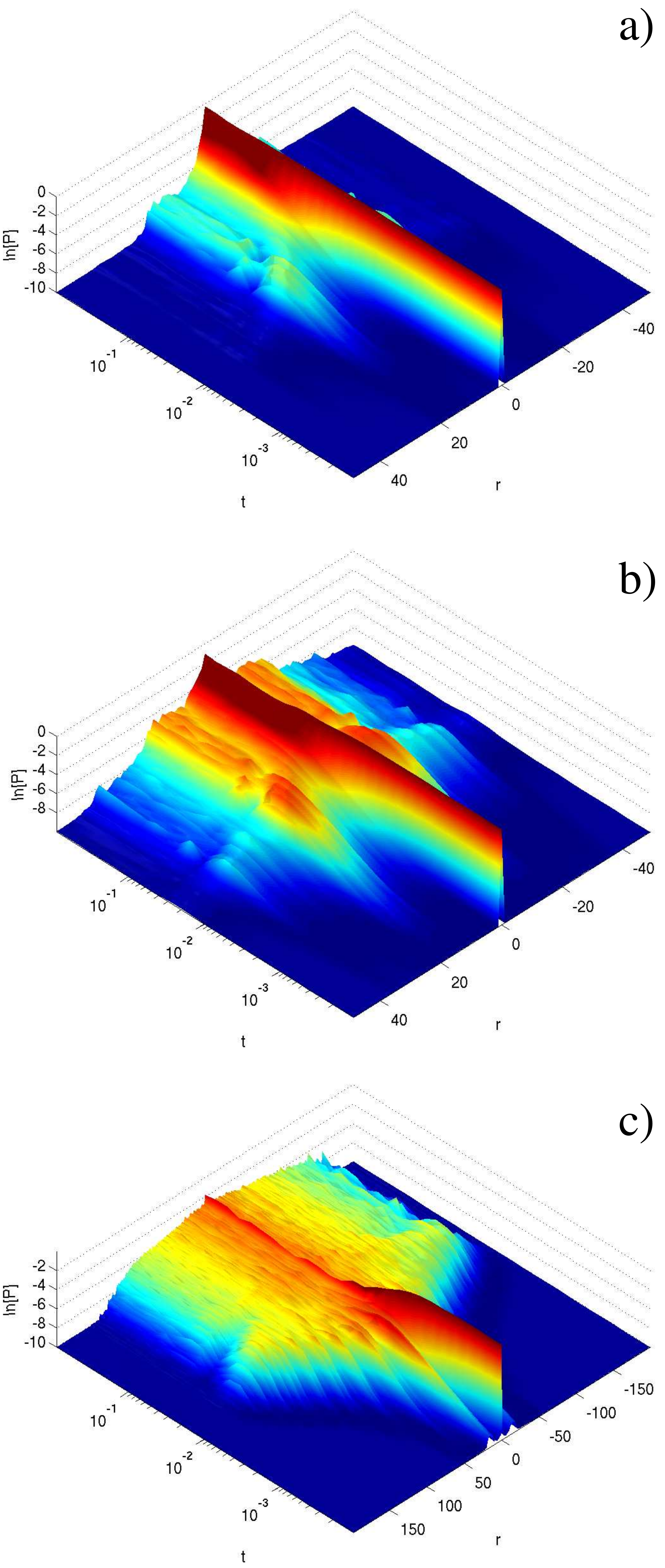}\hfill{}
  \vspace*{0.1cm}
\caption{\label{cha4-fig:P_qm_3D}  (Color online) The profile $P_{t}(r)$ of the BHH plotted
as a function of time for various perturbation strengths $\delta k<\delta k_{\tbox{qm}}$
(a), $\delta k_{\tbox{qm}}<\delta k<\delta k_{\tbox{prt}}$ (b), $\delta k>\delta k_{\tbox{prt}}$
(c). Note the different scale in (c). Here, $N=70$, $\tilde{E}=0.26$
and $\lambda_{0}=0.053$. }
\end{figure}

An overview of the spreading profiles for three representative $\delta k$-strengths is given 
in Fig.~\ref{cha4-fig:P_qm_3D}. A qualitative difference in the spreading is evident: In 
Fig.~\ref{cha4-fig:P_qm_3D}a the probability is mainly concentrated in the initial level for all 
times (standard perturbative regime). In Fig.~\ref{cha4-fig:P_qm_3D}b one can distinguish two 
different components in the $P_t(n|m)$, the {}``core" (characterized by $\delta E_{\rm core}(t)$) 
and the {}``tail" component (characterized by $\delta E(t)$), both of them being smaller than the 
bandwidth (extended perturbative regime). For even stronger perturbations, the core spills all 
over the bandwidth (see Fig.~\ref{cha4-fig:P_qm_3D}c) and the dynamics is non-perturbative. In 
the following we discuss each of these regimes separately.

%=====================================================================================
\subsubsection{The perturbative regimes}

For small perturbations $\delta k<\delta k_{\tbox{qm}}$ (see Fig.~\ref{cha4-fig:P_qm_3D}a) the 
probability is mainly concentrated in the initial level during the \emph{entire} evolution. This 
is the FOPT (standard perturbative) regime where the perturbation mixes only nearby levels 
and little probability escapes to the tails. 

\begin{figure*}[!t]
\includegraphics[%
  width=0.45\textwidth,
  keepaspectratio]{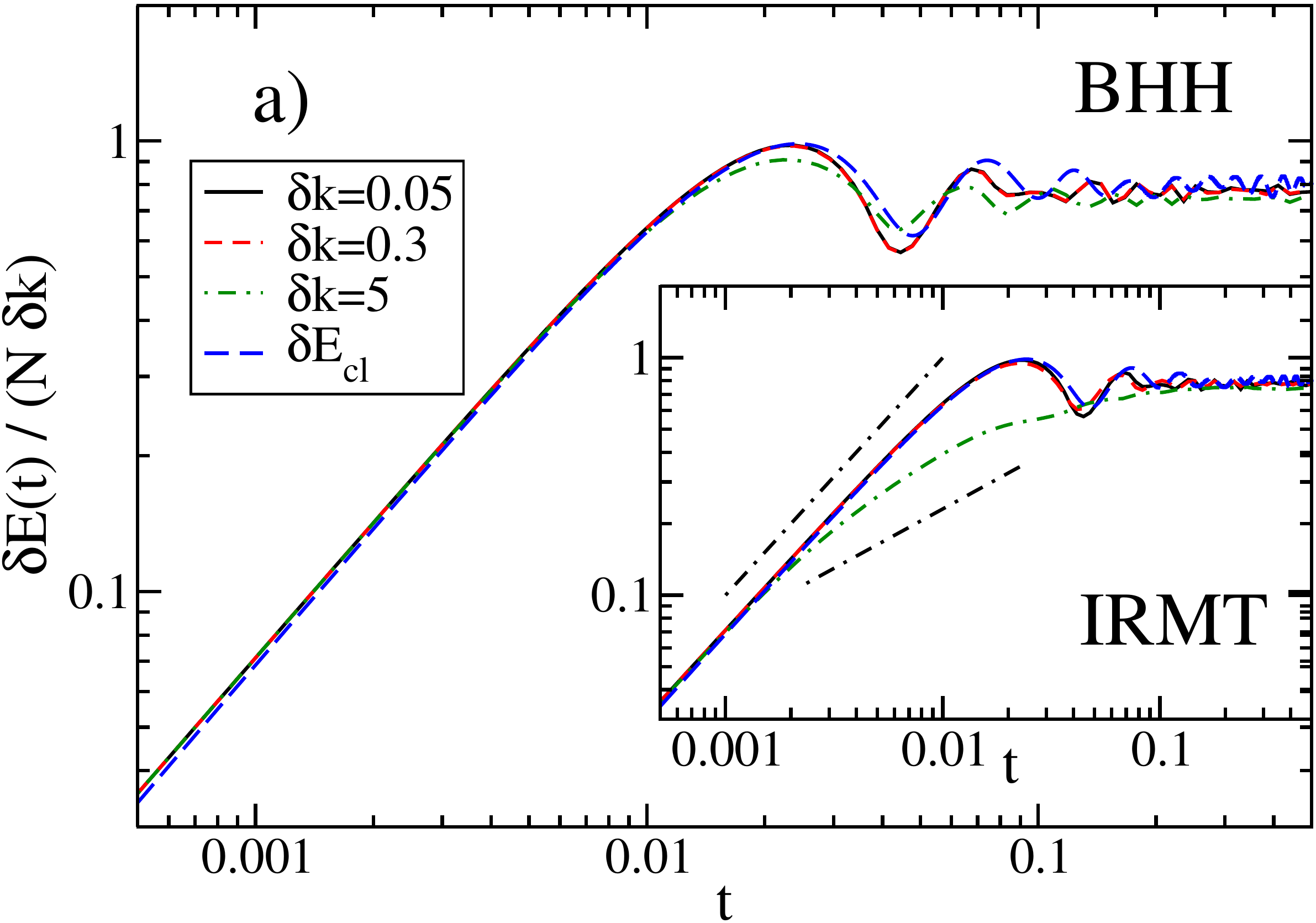}\hfill{}\includegraphics[%
  width=0.45\textwidth,
  keepaspectratio]{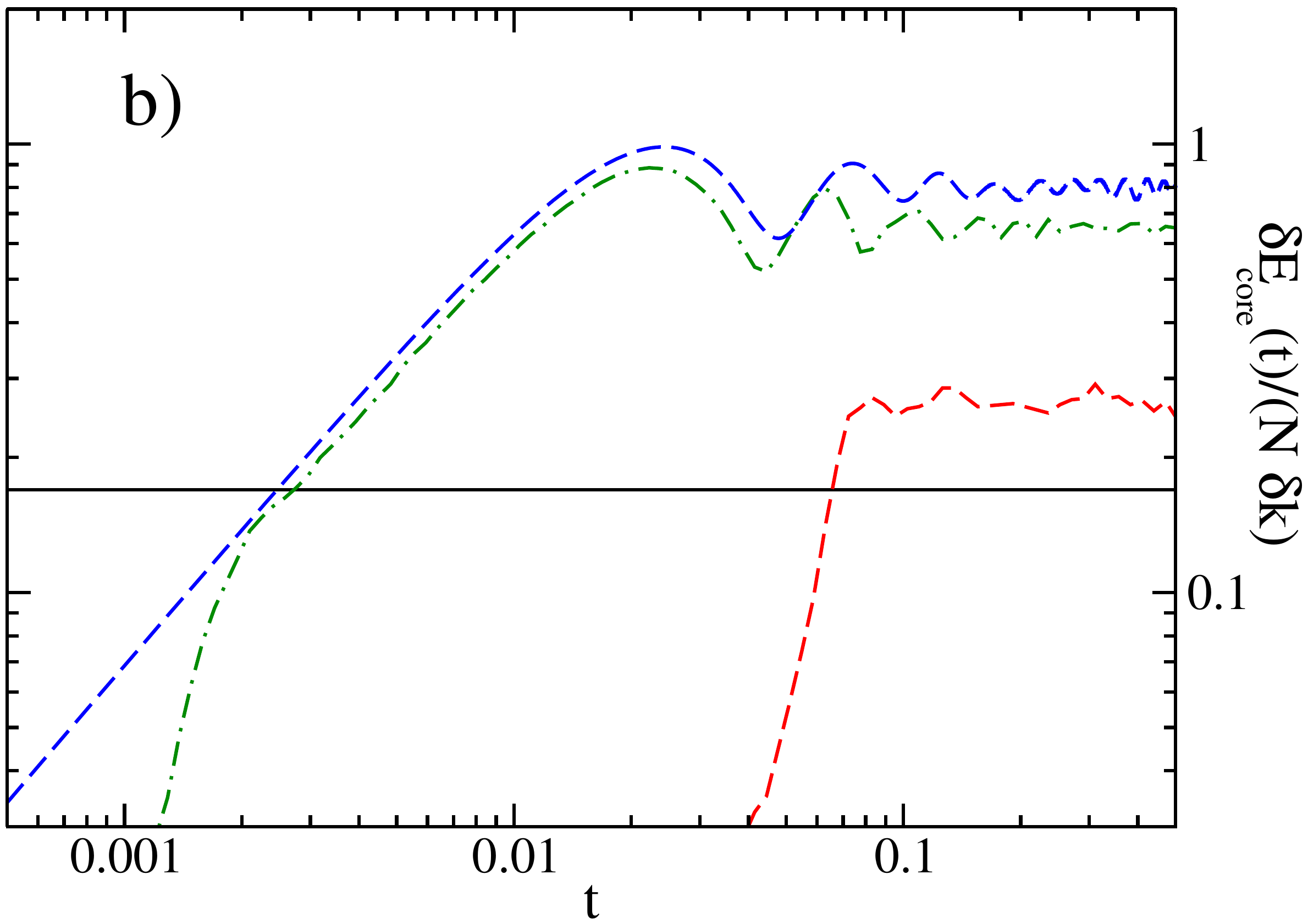}
\caption{\label{cha4-fig:dE_BHH_IRMT}(Color online){\it Left panel}: The (normalized) energy spreading $\delta E(t)$
for the BHH and the IRMT model (inset) for three different perturbation
strengths $\delta k\!=\!0.05\!<\!\delta k_{\tbox{qm}}$ (solid black line),
$\delta k_{\tbox{qm}}\!<\!\delta k\!=\!0.3<\delta k_{\tbox{prt}}$ (dashed
red line), and $\delta k=5>\delta k_{\tbox{prt}}$ (dash-dotted green
line). The classical expectation $\delta E_{\tbox{cl}}(t)$ is represented
in all three plots by a dashed blue line for comparison. In the inset
the black dash-dotted lines have slope one and one-half respectively
and are drawn to guide the eye. While for the BHH model one observes
restricted quantum-classical correspondence in all regimes this is
not the case for the IRMT model (inset): For perturbations $\delta k>\delta k_{\tbox{prt}}$
the energy spreading $\delta E(t)$ exhibits a premature crossover
to diffusive behavior. {\it Right panel}: The evolution 
of the corresponding core width $\delta E_{\tbox{core}}(t)$ for the BHH model is plotted.
In the perturbative regimes one observes a separation of scales 
$\delta E_{\tbox{core}}(t)<\delta E(t)<\Delta_{b}$,
which is lost for strong perturbations $\delta k>\delta k_{\tbox{prt}}$,
where $\delta E_{\tbox{core}}(t)$ approaches more and more the classical
expectation $\delta E_{\tbox{cl}}(t)$.
Here, $N=230$, $\tilde{U}=280$, $\tilde{E}=0.26$
and $\lambda_{0}=0.053$.
}
\end{figure*}

As the perturbation strength is increased $\delta k_{\tbox{qm}} <\delta k <\delta k_{\tbox{prt}}$ 
(Fig.~\ref{cha4-fig:P_qm_3D}b), levels within the bandwidth are mixed and one can distinguish two 
different components in the profile $P_{t}(r)$: The core characterized by $\delta E_{\tbox{core}}
(t)$, where most of the probability is concentrated, and the tail component, characterized by 
$\delta E(t)$. The latter is reported in Fig.~\ref{cha4-fig:dE_BHH_IRMT}a together with the classical 
spreading $\delta E_{\tbox{cl}}(t)$. The remarkable fact is that, as far as $\delta E(t)$ is 
concerned, the agreement with the classical result is perfect. This might lead to the wrong impression 
that the classical and quantum spreading are of the same nature. However, this is definitely not 
the case.

In order to reveal the different nature of the quantum spreading in the perturbative regime we 
turn to the analysis of the core-width $\delta E_{\tbox{core}}(t)$ (see Fig.~\ref{cha4-fig:dE_BHH_IRMT}b). 
If the spreading were of classical type this would imply that the evolving profile would be 
characterized by a single energy scale, and thus $\delta E(t)\sim\delta E_{\tbox{core}}(t)$. 
However, as can be seen in Fig.~\ref{cha4-fig:dE_BHH_IRMT}b this is certainly not the case:
For $\delta k<\delta k_{\tbox{qm}}$ we have that $\delta E_{\rm core}(t)=\Delta$ for all times
while for $\delta k_{\tbox{qm}} <\delta k <\delta k_{\tbox{prt}}$ the core-width fulfills
the inequalities $\Delta < \delta E_{\rm core}(t)<\delta E(t)<\Delta_b$. In fact, this separation
of energy scales allows us to use perturbation theory in order to evaluate theoretically
the evolving second moment of the energy distribution. We get for the transition probability 
from an initial state $m$ to any other state $n\neq m$ 
\begin{equation}
P_{t}(n|m) = 
\frac{\delta k^{2}}{\hbar^2}|\bm{B}_{nm}|^{2}
\frac{ \tilde{F}_{t}(\omega_{nm}) } { (\omega_{nm})^2 }
\label{eq:FOPT-ker}
\end{equation}
Here $\tilde{F}_t(\omega)=(\omega t)^2{\cdot}\mbox{sinc}^2(\omega t/2)$ is the spectral content of a constant 
perturbation of duration $t$, and $\mbox{sinc}(x)=\sin(x)/x$. Substituting the above expression in
Eq.~(\ref{cha4-eq:dE_t}) we get the LRT expression (\ref{eq:LRT-dE-wpk2}) for $\delta E(t)$. We 
have also calculated the second moment resulting from the IRMT modeling. The outcome is reported in 
the inset of Fig.~\ref{cha4-fig:dE_BHH_IRMT}a and shows that within the perturbative regime the IRMT 
modeling provides the same results (as far as the second moment is concerned) as the LRT calculations. 
Therefore we conclude that for $\delta k \le \delta k_{\rm prt}$ the IRMT modeling, the LRT results, 
the classical results $\delta E_{\rm cl}$, and the quantum calculations for the second moment $\delta 
E(t)$ of the BHH match one another.

Encouraged by this success of LRT and the IRMT modeling to describe the second moment $\delta E (t)$ of the 
energy spreading, we can further use them to evaluate the survival probability ${\cal P}(t)$. Assuming 
a Markovian picture of the dynamics, LRT predicts  \cite{HCGK06}
\begin{eqnarray}
{\cal P}(t) = \exp\left[
-\delta k^2 \times
\int_{-\infty}^{\infty}
\frac{d{\omega}}{2\pi}
\tilde{C}(\omega)
\frac{\tilde{F}_{t}(\omega)}{(\hbar\omega)^{2}}\,.
\right]
\label{eq:LRT-qm-P}
\end{eqnarray}
which after substituting the spectral-content ${\tilde F}_t(\omega)$, can be re-written in the following 
form
\begin{equation}
\mcal{P}(t)=\exp\left[-\left(\frac{\delta k}{\hbar}\right)^2\times\int_{-\infty}^{\infty}\frac{d{\omega}}{2\pi}\tilde{C}(\omega)\,\, 
t\,\,\, t\,\mbox{sinc}^{2}\left(\frac{\omega t}{2}\right)\right]\szp
\label{cha4-eq:P_LRT_inwpdsection}
\end{equation}
For short times ($t\ll\tau_{\tbox{cl}}$) during which the spreading is ballistic-like, the term 
$t\,\mbox{sinc}^{2}(\omega t/2)$ is broad compared to the band profile and can be approximated by $t$ 
leading to 
\begin{equation}
{\mathcal{P}}(t)={\textrm{exp}}\left[-C(\tau{=}0)\times\left({\frac{\delta k\, t}{\hbar}}\right)^{2}\right]\szp
\label{cha4-eq:P_prt_short_times}
\end{equation}
For longer times ($t\gg\tau_{\tbox{cl}}$) on the other hand, the term $t\,\mbox{sinc}^{2}(\omega t/2)$ is 
extremely narrow and can be approximated by a delta function $\delta(\omega)$. This results in a 
Fermi-Golden-Rule (FGR) decay 
\begin{equation}
\mathcal{P}(t)=\exp\left[-\left(\frac{\delta k}{\hbar}\right)^{2}\tilde{C}(\omega=0)\times t\right]\szk
\label{cha4-eq:P_prt_FGR}
\end{equation}
which can be trusted as long as $\mathcal{P}(t)\sim1$. This can be converted into an inequality
$t<t_{\tbox{prt}} =\left(\frac{\delta k_{\tbox{prt}}}{\delta k}\right)^{2}\tau_{\tbox{cl}}$. 

\begin{figure*}[!t]
\includegraphics[%
  width=0.45\textwidth,
  keepaspectratio]{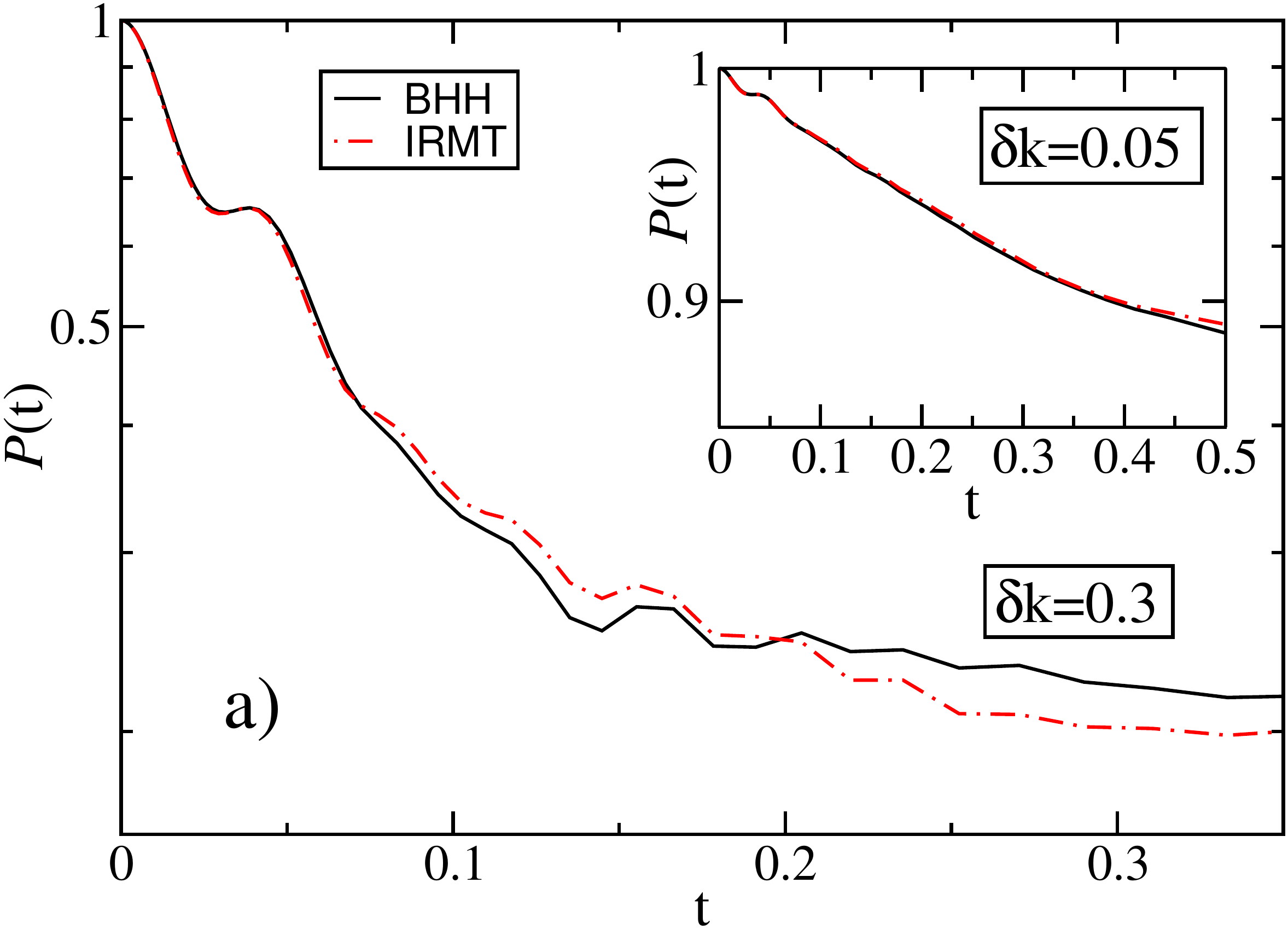}\hfill{}\includegraphics[%
  width=0.45\textwidth,
  keepaspectratio]{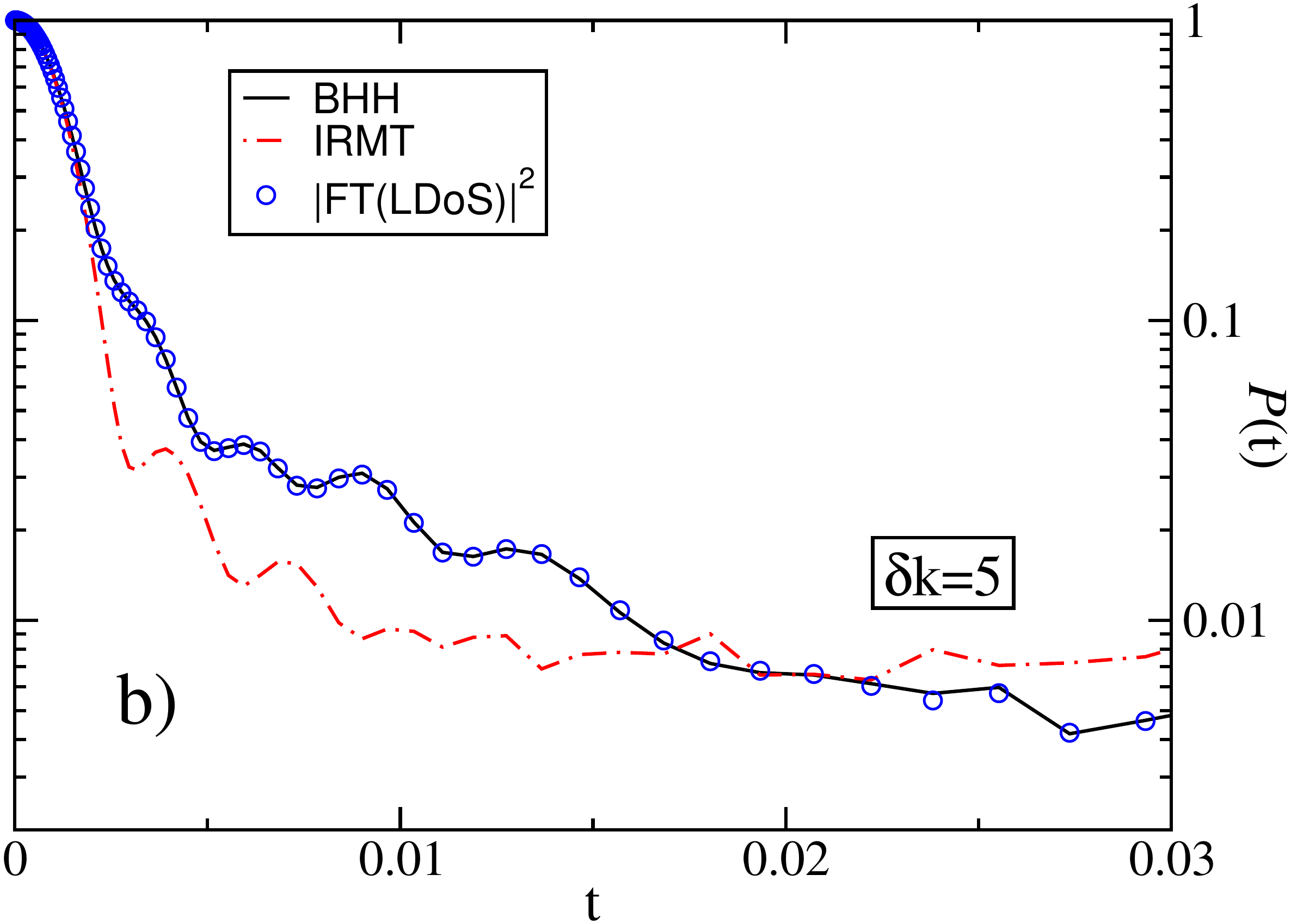}
\caption{\label{cha4-fig:P_IRMT}  (Color online) The survival probability $\mcal{P}(t)$ for
the BHH and three different perturbation strengths a)$\delta k=0.05<\delta k_{\tbox{qm}}$
(inset), $\delta k_{\tbox{qm}}<\delta k=0.3<\delta k_{\tbox{prt}}$
(main figure), and b) $\delta k=5>\delta k_{\tbox{prt}}$ . The solid
black line represent the exact numerical result while the dash-dotted
red line is the LRT result (\ref{eq:LRT-qm-P}) calculated using
the IRMT model. The
inset of panel (a) represents the FOPT regime, while the main figure
corresponds to the extended perturbative regime. Here the break time
is $t_{\tbox{prt}}\sim0.1$ (see Subsection~(\ref{sec:nonprt-WPD}).
In the non-perturbative regime (b), the LRT breaks down close to the
calculated break time $t_{\tbox{prt}}\sim0.001$. In this panel we
superimpose the Fourier transform of the LDoS as blue circles. The
agreement with $\mcal{P}(t)$ is excellent. Here, $N=230$, $\tilde{U}=280$,
$\tilde{E}=0.26$ and $\lambda_{0}=0.053$.}
\end{figure*}

In Fig.~\ref{cha4-fig:P_IRMT}a we plot our numerical results for the trimeric BHH model together with 
the theoretical expectation (\ref{eq:LRT-qm-P}) (we note that the outcome of the IRMT modeling matches
exactly the results of the LRT and thus we do not overplot them). 
In both perturbative regimes we observe a short initial Gaussian decay
(as implied by Eq.~(\ref{cha4-eq:P_prt_short_times})) which is followed
by the exponential FGR decay. In the FOPT regime (inset of Fig.~\ref{cha4-fig:P_IRMT}a)
the entire decay until saturation is described by LRT. In the extended
perturbative regime (see Fig.~\ref{cha4-fig:P_IRMT}a) the overall
agreement is still pretty good. However, here the perturbative break
time $t_{\tbox{prt}}$ is shorter and one finds a deviation around
the time $t_{\tbox{prt}}\sim0.01$.

%=====================================================================================
\subsubsection{The non-perturbative regime\label{sec:nonprt-WPD}}

Once we enter the non-perturbative regime $\delta k>\delta k_{\tbox{prt}}$ (see Fig.~\ref{cha4-fig:P_qm_3D}c), 
the core spills over the bandwidth and the separation of energy scales is lost, leading to $\delta E(t) \sim 
\delta E_{\tbox{core}}(t) > \Delta _b$ (see Fig.~\ref{cha4-fig:dE_BHH_IRMT}b for $\delta k =5$). In this
case the evolving energy distribution becomes totally non-perturbative. Still, for short times $t_{\tbox{prt}}
=\left(\frac{\delta k_{\tbox{prt}}}{\delta k}\right)\tau_{\tbox{cl}}< \tau_{\rm cl}$, defined by the requirement 
that ${\cal P}(t)\sim 1$ (see Eq. (\ref{cha4-eq:P_prt_short_times})), the evolving probability kernel $P_t(n|m)$ 
(and therefore the spreading $\delta E(t)$) is described accurately by the FOPT expression (\ref{eq:FOPT-ker}).

The remarkable fact is that although for $t>t_{\rm prt}$ the evolving profile $P(n|m)$ is totally non-perturbative,
this crossover is {\em not} reflected in the variance (see Fig.~\ref{cha4-fig:dE_BHH_IRMT}a). The agreement 
with the LRT results of Eq.(\ref{eq:LRT-dE-wpk2}) is still perfect. Instead, the crossover can be detected by studying 
other moments like $\delta E_{\rm core}(t)$ which acquire classical characteristics, i.e. $\delta E_{\rm core}(t)
\approx \delta E(t)= \delta E_{\rm cl}(t)$ (see Fig.~\ref{cha4-fig:dE_BHH_IRMT}b). Thus we are led to the 
conclusion \cite{Cohe00,KC01} that the LRT predictions are not applicable while detailed QCC would possibly 
validate semiclassical considerations. We will examine this assumption more carefully in Subsection 
\ref{cha4-sec:WPD-detail}. 

What about the IRMT modeling? In the inset of Fig.~\ref{cha4-fig:dE_BHH_IRMT}a we report the numerical 
results for the spreading $\delta E(t)$ of the IRMT model. We observe that as soon as we enter the
non-perturbative regime, the spreading $\delta E(t)$ shows a qualitatively different behavior than the 
dynamical BHH model. Namely, after an initial ballistic spreading (taking place for times $t<t_{\rm prt}$), 
we observe a premature crossover to a diffusive behavior $\delta E(t)=\sqrt{2D_{E}t}$. The following 
heuristic picture can explain the diffusive behavior of the IRMT modeling. At $t\sim t_{\tbox{prt}}\ll 
\tau_{\rm cl}$, the evolving distribution becomes as wide as the bandwidth, and we have $\delta E_{
\tbox{core}}\sim\delta E\sim \Delta_{b}$ rather than $\delta E_{\tbox{core}}\ll\delta E\ll\Delta_{b}$. 
Once the mechanism for ballistic-like spreading disappears, a stochastic-like behavior takes its place. 
This is similar to a random-walk process where the step size is of the order $\Delta_{b}$, with transient 
time $t_{\tbox{prt}}$. 

\begin{figure*}
\hfill{}\includegraphics[%
  width=\textwidth,
  keepaspectratio]{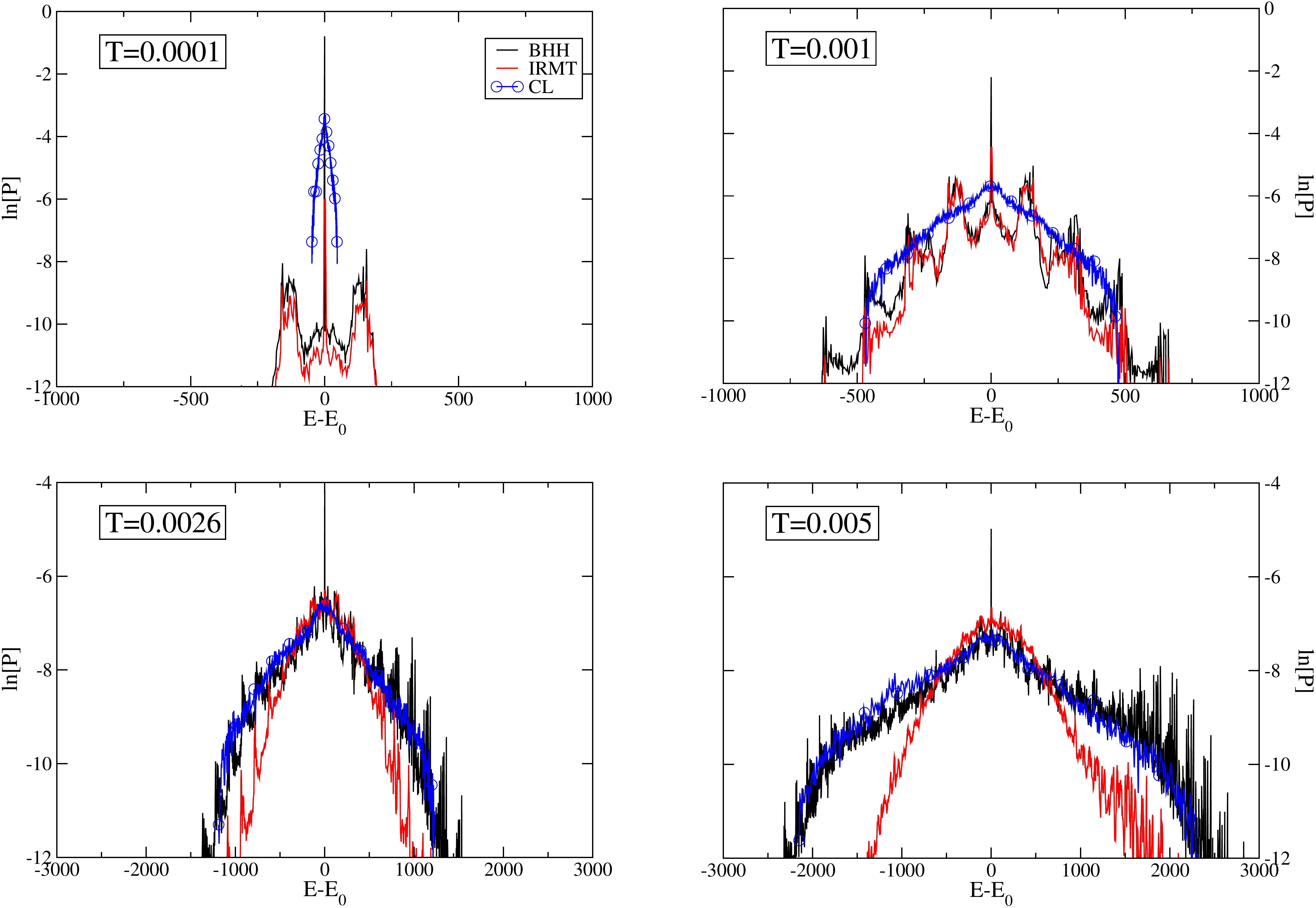}\hfill{}
\caption{\label{cha4-fig:P_snapshots}   (Color online) Snapshots of the evolving quantum profile
$P_{t}(r)$ obtained from the BHH (black line) and the IRMT model
(red line) as well as the classical profile $P_{t}^{\tbox{cl}}(r)$
(blue line with $\circ$) in the non-perturbative regime $\delta k=5>\delta k_{\tbox{prt}}$
plotted against the energy difference $E-E_{0}$. After the quantal
transition period $t\sim0.002$ (see Fig.~\ref{cha4-fig:dE_BHH_IRMT}b)
there is no scale separation between the core and the tail component
and one observes overall detailed QCC. However, the initially excited
component $\tket{n_{0}}$ decays slower in the quantum case. Here,
$N=230$, $\tilde{U}=280$, $\tilde{E}=0.26$ and $\lambda_{0}=0.053$.}
\end{figure*}

The same deviations are observed for other observables as well. In Fig. \ref{cha4-fig:P_IRMT}b we 
report our results for the survival probability in the non-perturbative regime. We find that the 
IRMT modeling (which for short times gives the same results as LRT--not shown in the figure as they 
are indistinguishable from the IRMT results) breaks down after an initial Gaussian decay (\ref{cha4-eq:P_prt_short_times}) 
which holds up to a break time $t_{\tbox{prt}}\sim0.001$. Instead, the behavior of $\mathcal{P}(t)$ 
can be obtained by a Fourier transform of the LDoS. Specifically, we have that
\begin{eqnarray}
  \mathcal{P}(t) & \equiv & \left|\langle n(k_{0})|e^{-i\hat{H}(k)t/\hbar}|n(k_{0})\rangle\right|^{2}\nonumber\\
  & = & \left|\sum_{m}e^{-iE_{m}(k)t/\hbar}|\langle m(k)|n(k_{0})\rangle|^{2}\right|^{2}\nonumber \\
  & = & \left|\int_{\infty}^{\infty}P(E|m)e^{-iEt/\hbar}dE\right|^{2}\szk
  \label{cha4-eq:P_is_FT_of_LDOS}
\end{eqnarray}
where $P(E|m)$ is given by Eq.~(\ref{LDoS}). In Fig.~\ref{cha4-fig:P_IRMT}b we superimpose the outcome 
of Eq. (\ref{cha4-eq:P_is_FT_of_LDOS}) (see blue circles) together with the survival probability evaluated
by the numerical integration of the Schr\"odinger equation. An excellent agreement is evident.

%=====================================================================================
\subsection{Detailed versus restricted QCC \label{cha4-sec:WPD-detail}}

In the previous subsection we have assumed that the evolving wavepacket is developing detailed QCC 
in the non-perturbative regime and for times $t> t_{\rm prt}=\left(\frac{\delta k_{\tbox{prt}}}{\delta 
k}\right)\tau_{\tbox{cl}}$ (for earlier times FOPT --or equivalently IRMT considerations-- apply). 

In Fig.~\ref{cha4-fig:P_snapshots} we report four snapshots of the evolving quantum mechanical profile
(black lines). In the same figure we report the IRMT results (red lines) together with the classical 
calculations (blue lines with $\circ$). As we have discussed above we distinguish two phases in the evolution: 
For $t<\tau_{\rm cl}$ the IRMT modeling (or equivalently the FOPT) is applicable while for $t>\tau_{\rm cl}$
the evolving profile is described by its classical counterpart $P_{\rm cl}(t)$. During this second
phase, the evolution predicted by the IRMT is diffusive leading to a Gaussian shape for $P_t(n|m)$.

%%%%%%%%%%     CONCLUSIONS    %%%%%%%%%%
\section{Conclusions \label{sec:conclusions}}

In this paper we have studied the evolving energy distribution of a three-site ring-shaped
Bose-Hubbard model in the chaotic regime. The evolution is triggered by a change $\delta k$
in the tunneling rate $k$ between neighboring lattice sites which in the context of ultra-cold
atoms in optical lattices is realized by a change in the intensity of the trapping laser field.
The specific scenario that we have analyzed in detail is the so-called wavepacket dynamics
in energy space corresponding to a constant driving pulse of finite duration $t$. 

We followed a three-fold approach to the problem which combines purely quantum mechanical
as well as semiclassical and random matrix theory considerations. This enabled us to identify
both the strengths and limitations of each method.

We find the appearance of three dynamical $\delta k$-regimes: The standard perturbative
($\delta k<\delta k_{\tbox{qm}}\propto\tilde{U}/N^{3/2}$), the extended perturbative
($\delta k_{\tbox{qm}}<\delta k<\delta k_{\tbox{prt}}\propto\tilde{U}/{N}$) and the
non-perturbative regime ($\delta> \delta k_{\tbox{prt}}$). The first two regimes can be
addressed using LRT or RMT calculations. In contrast, the last regime requires a combination
of LRT/RMT calculations and semiclassical considerations. The former approach describes the
evolving energy distribution for short times while the latter applies for longer times. Interestingly
enough we have found that the variance $\delta E^2(t)$ of the evolving energy distribution shows
a robust quantum-classical correspondence for all $\delta k$-values, while other moments exhibit
this QCC only in the non-perturbative regime identified with the classical limit. In this regime, even
an improved RMT modeling fails to describe the long time behavior of $\delta E(t)$ leading to a
premature crossover from ballistic to diffusive behavior.

The motivation of the present study is driven both by theoretical and experimental considerations.
On the fundamental level, we would like to understand the manifestation of quantum-classical
correspondence in the context of quantum chaotic dynamics, where chaos enters not due to geometrical
considerations ("chaotic" shape of the trap) but due to many-body interactions \cite{SHM02}.
At the same time, our results are also of immediate relevance to various branches of physics.
For example, in the framework of ultra-cold atoms loaded in optical traps one is interested in
understanding measurements of the energy absorption rates induced by potential modulations
\cite{SMSKE04,SSMKE04,KIGHS06,RBPWC05,ICHG06,BASD05,Lund04b}. Another application
arises in molecular physics: As mentioned in Section \ref{cha2-sec:BHH}, the Bose-Hubbard
Hamiltonian also models bond-excitations in small molecules \cite{SLE85,LSP94}. In this respect,
the wavepacket dynamics investigated here describes the vibrational energy redistribution of an
initial excitation \cite{LSP94}. 

As far as the experimental realization of our study is concerned, micro-traps \cite{Reich02}
are promising candidates for such time-dependent potentials \cite{Jo_etal07} while optical
lattices have already been successfully used in similar setups. Specifically, the studied dynamical
scenario is readily implemented by changing the intensity of the laser field using a simplified
version of the experiments of the Zurich group \cite{SMSKE04,SSMKE04}. In contrast to the
periodic modulation presented there, the optical lattice depth has to be altered in a step-like
manner. Such experiments have been successfully performed by Greiner et al. \cite{GMHB02}
where the intensity of the trapping laser field was suddenly raised. The raise time was achieved
to be much faster than the tunneling time between neighboring sites but slow enough as not
to excite higher vibrational modes of the wells. 

Concerning the measurement of the energy distribution $P_t(E)$ and the associated absorption of energy
due to the driving, various techniques may be applied. Using standard time-of-flight measurements one can
determine, for example, the release energy of the condensate and the momentum distribution of the atomic
cloud which we expect to provide the relevant information on the variance $\delta E^2(t)$ of the
energy distribution. Another possibility is to probe the $P_t(E)$ via phase diffusion measurements \cite{GMHB02}.
Experimentally, the BEC can be prepared (almost) in one eigenstate. The driving pulse induces a broadening
in the energy distribution leading to (decaying) oscillations in the contrast $\langle b_i^\dagger b_{i+1}+b_ib_{i+1}^\dagger\rangle$
between neighboring sites. We expect that the functional form of the decay can be directly related to the
core width $\Gamma$ and thus be used to detect the three parametric $\delta k$-regimes. 
While these measurements are in principle sensitive to decoherence due to residual interaction with the
non-condensed atoms, we note here that for two-site systems coherence times of several hundred
milliseconds were observed \footnote{M. Oberthaler, private communication.}

\begin{acknowledgments}
The authors acknowledge fruitful discussions with Doron Cohen, Sergei Flach, George Kalosakas,
and Markus Oberthaler. This research was supported by a grant from the United States-Israel
Binational Science Foundation (BSF) and the DFG within the Forschergruppe 760.
\end{acknowledgments}

%%%%%%%%%%%%%%%%%%%%%%%%%%%%%%%%%%%%%%%%%%%

\bibliography{paper,books}
%\bibliographystyle{./apsrev_nonotes}

%%%%%%%%%%%%%%%%%%%%%%%%%%%%%%%%%%%%%%%%%%%

\end{document}